\documentclass{aastex}
\usepackage{emulateapj5}
\usepackage{apjfonts}
\usepackage{epsf}
\slugcomment{accepted for publication in ApJ} 
\shorttitle{SDSS J1004+4112}
\shortauthors{OGURI ET AL.}
\begin{document}
%%%%%%%%%%%%%%%%%%%%%%%%%%%%%%%%%%%%%%%%%%%%%%%%%%%%%%%%%%%%%%%%%%%%%%
%%%%%%%%%%%%%%%%%%%%%%%%%%%%%%%%%%%%%%%%%%%%%%%%%%%%%%%%%%%%%%%%%%%%%%
\title{Observations and Theoretical Implications of the Large Separation
Lensed Quasar SDSS J1004+4112} 
%%%%%%%%%%%%%%%%%%%%%%%%%%%%%%%%%%%%%%%%%%%%%%%%%%%%%%%%%%%%%%%%%%%%%%
%%%%%%%%%%%%%%%%%%%%%%%%%%%%%%%%%%%%%%%%%%%%%%%%%%%%%%%%%%%%%%%%%%%%%%
%
%%%%%%%%%%%%%%%%%%%%%%%%%%%%%%%%%%%%%%%%%%%%%%%%%%%%%%%%%%%%%%%%%%%%%%
\author{Masamune Oguri,\altaffilmark{1}
Naohisa Inada,\altaffilmark{1}
Charles R. Keeton,\altaffilmark{2}
Bartosz Pindor,\altaffilmark{3}\\
Joseph F. Hennawi,\altaffilmark{3}
Michael D. Gregg,\altaffilmark{4,5}
Robert H. Becker,\altaffilmark{4,5}
Kuenley Chiu,\altaffilmark{6}\\
Wei Zheng,\altaffilmark{6}
Shin-Ichi Ichikawa,\altaffilmark{7}
Yasushi Suto,\altaffilmark{1}
Edwin L. Turner,\altaffilmark{3}\\
James Annis,\altaffilmark{8}
Neta A. Bahcall,\altaffilmark{3}
Jonathan Brinkmann,\altaffilmark{9}
Francisco J. Castander,\altaffilmark{10}\\
Daniel J. Eisenstein,\altaffilmark{11}
Joshua A. Frieman,\altaffilmark{2,8}
Tomotsugu Goto,\altaffilmark{6,12,13}
James E. Gunn,\altaffilmark{3}\\
David E. Johnston,\altaffilmark{2}
Stephen M. Kent,\altaffilmark{8}
Robert C. Nichol,\altaffilmark{14}
Gordon T. Richards,\altaffilmark{3}\\
Hans-Walter Rix,\altaffilmark{15}
Donald P. Schneider,\altaffilmark{16}
Erin Scott Sheldon,\altaffilmark{2}\\
and
Alexander S. Szalay\altaffilmark{6}
}
\altaffiltext{1}{Department of Physics, University of Tokyo, Hongo 7-3-1,
Bunkyo-ku, Tokyo 113-0033, Japan.}
\altaffiltext{2}{Astronomy and Astrophysics Department, University of
Chicago, 5640 South Ellis Avenue, Chicago, IL 60637.}
\altaffiltext{3}{Princeton University Observatory, Peyton Hall,
Princeton, NJ 08544.}
\altaffiltext{4}{Department of Physics, University of California at
Davis, 1 Shields Avenue, Davis, CA 95616.}
\altaffiltext{5}{Institute of Geophysics and Planetary Physics,
Lawrence Livermore National Laboratory, L-413, 7000 East Aveneu,
Livermore, CA 94550.}
\altaffiltext{6}{Department of Physics and Astronomy, Johns Hopkins
University, 3701, San Martin Drive, Baltimore, MD 21218.}
\altaffiltext{7}{National Astronomical Observatory, 2-21-1 Osawa,
Mitaka, Tokyo 181-8588, Japan.}
\altaffiltext{8}{Fermi National Accelerator Laboratory, P.O. Box
500, Batavia, IL 60510.}
\altaffiltext{9}{Apache Point Observatory, P.O. Box 59, Sunspot, NM88349.}
\altaffiltext{10}{Institut d'Estudis Espacials de Catalunya/CSIC,
Gran Capita 2-4, 08034 Barcelona, Spain.}
\altaffiltext{11}{Steward Observatory, University of Arizona,
933 North Cherry Avenue, Tucson, AZ 85721.}
\altaffiltext{12}{Department of Astronomy, University of Tokyo, Hongo
7-3-1, Bunkyo-ku, Tokyo 113-0033, Japan.}
\altaffiltext{13}{Institute for Cosmic Ray Research, University of Tokyo, 
5-1-5 Kashiwa, Kashiwa City, Chiba 277-8582, Japan.}
\altaffiltext{14}{Department of Physics, Carnegie Mellon University,
Pittsburgh, PA 15213.}
\altaffiltext{15}{Max-Planck Institute for Astronomy, K\"onigstuhl
17, D-69117 Heidelberg, Germany.}
\altaffiltext{16}{Department of Astronomy and Astrophysics, Pennsylvania State
University, 525 Davey Laboratory, University Park, PA 16802.}
%%%%%%%%%%%%%%%%%%%%%%%%%%%%%%%%%%%%%%%%%%%%%%%%%%%%%%%%%%%%%%%%%%%%%%
%
%\received{}
%\accepted{}
%
\begin{abstract}
We study the recently discovered gravitational lens SDSS~J1004+4112,
the first quasar lensed by a cluster of galaxies. It consists of four
images with a maximum separation of $14\farcs62$. The system was
selected from the photometric data of the Sloan Digital Sky Survey
(SDSS), and has been confirmed as a lensed quasar at $z=1.734$ on the
basis of deep imaging and spectroscopic follow-up observations. We
present color-magnitude relations for galaxies near the lens plus
spectroscopy of three central cluster members, which unambiguously
confirm that a cluster at $z=0.68$ is responsible for the large image
separation. We find a wide range of lens models consistent with the
data, and despite considerable diversity they suggest four general
conclusions: (1) the brightest cluster galaxy and the center of the
cluster potential well appear to be offset by several kiloparsecs;
(2) the cluster mass distribution must be elongated in the North--South
direction, which is consistent with the observed distribution of cluster
galaxies; (3) the inference of a large tidal shear ($\sim$0.2) suggests
significant substructure in the cluster; and (4) enormous uncertainty in
the predicted time delays between the images means that measuring the
delays would greatly improve constraints on the models. We also compute
the probability of such large separation lensing in the SDSS quasar
sample, on the basis of the Cold Dark Matter model. The lack of large
separation lenses in previous surveys and the discovery of one in SDSS
together imply a mass fluctuation normalization
$\sigma_8=1.0^{+0.4}_{-0.2}$ (95\% confidence), if cluster dark matter 
halos have an inner density profile $\rho \propto r^{-1.5}$. Shallower
profiles would require higher values of $\sigma_8$. Although the
statistical conclusion might be somewhat dependent on the degree of
the complexity of the lens potential, the discovery of SDSS~J1004+4112
is consistent with the predictions of the abundance of cluster-scale
halos in the Cold Dark Matter scenario.
\end{abstract}
 
 
%%%%%%%%%%%%%%%%%%%%%%%%%%%%%%%%%%%%%%%%%%%%%%%%%%%%%%%%%%%
%%%%%%%%%%%%%%%%%%%%%%%%%%%%%%%%%%%%%%%%%%%%%%%%%%%%%%%%%%%
\section{Introduction}
%%%%%%%%%%%%%%%%%%%%%%%%%%%%%%%%%%%%%%%%%%%%%%%%%%%%%%%%%%%
%%%%%%%%%%%%%%%%%%%%%%%%%%%%%%%%%%%%%%%%%%%%%%%%%%%%%%%%%%%

Since the discovery of the first gravitationally lensed quasar Q0957+561
\citep*{walsh79}, about 80 strong lens systems have been found so far.
All of the lensed quasars have image separations smaller than $7''$,
and they are lensed by massive galaxies (sometimes with small boosts
from surrounding groups or clusters of galaxies). The probability that
distant quasars are lensed by intervening galaxies was originally
estimated by \citet*{turner84} to be 0.1\%--1\%, assuming that galaxies
can be modeled as singular isothermal spheres (SIS). This prediction
has been verified by several optical and radio lens surveys, such as the
{\it Hubble Space Telescope (HST)} Snapshot Survey \citep{bahcall92},
the Jodrell Bank/Very Large Array Astrometric Survey
\citep[JVAS;][]{patnaik92}, and the Cosmic Lens All Sky Survey
\citep[CLASS;][]{myers95}.  The lensing probability is sensitive to the
volume of the universe, so it can be used to place interesting
constraints on the cosmological constant $\Omega_\Lambda$ \citep*[][but
see \citealt{keeton02}]{turner90,fukugita90,kochanek96,chiba99,chae02}.

In contrast, lenses with larger image separations should probe a
different deflector population: massive dark matter halos that host
groups and clusters of galaxies. Such lenses therefore offer valuable
and complementary information on structure formation in the universe,
including tests of the Cold Dark Matter (CDM) paradigm
\citep{narayan88,cen94,wambsganss95,kochanek95a,flores96,nakamura97}.
So far the observed lack of large separation lensed quasars has been
used to infer that, unlike galaxies, cluster-scale halos cannot be
modeled as singular isothermal spheres \citep*{keeton98a,porciani00,
kochanek01,keeton01a,sarbu01,li02,li03,oguri02b,ma03}.  The difference
can probably be ascribed to baryonic processes: baryonic infall and
cooling have significantly modified the total mass distribution in
galaxies but not in clusters \citep{rees77,blumenthal86, kochanek01}.
As a result, large separation lenses may constrain the density profiles
of dark matter halos of cluster more directly than small separation
lenses \citep*{maoz97, keeton01c,wyithe01,takahashi01,li02,oguri02a,
oguri03a, huterer04,kuhlen04}.
Alternatively, large separation lensed quasars may be used to place
limits on the abundance of massive halos if the density profiles are
specified \citep{narayan88,wambsganss95,kochanek95a, nakamura97,
mortlock00,oguri03a,lopes04}.
Better yet, the full distribution of lens image separations may provide
a systematic diagnostic of baryonic effects from small to large scales
in the CDM scenario.

The fact that clusters have less concentrated mass distributions than
galaxies implies that large separation lensed quasars should be less
abundant than small separation lensed quasars by one or two orders of
magnitude. This explains why past surveys have failed to unambiguously
identify large separation lensed quasars
\citep*{kochanek95b,phillips01a,zhdanov01,ofek01,ofek02}.  For instance,
CLASS found 22 small separation lenses but no large separation lenses
among $\sim$15000 radio sources \citep{phillips01b}.  Although several
large separation lensed quasar candidates have been found
\citep*[e.g.,][]{mortlock99}, they are thought to be physical (unlensed)
pairs on the basis of individual observations \citep[e.g.,][]{green02}
or statistical arguments \citep*{kochanek99,rusin02}. Recently
\citet{miller04} found 6 candidate lens systems with image separations
$\theta > 30''$ among $\sim$20000 quasars in the Two-degree Field (2dF)
quasar sample.  Given the lack of high-resolution spectra and deep
imaging for the systems, however, it seems premature to conclude that
they are true lens systems. We note that because the expected number
of lenses with such large image separations in the 2dF sample is much
less than unity \citep{oguri03a}, these systems would present a severe
challenge to standard models if confirmed as lenses.

To find a first unambiguous large separation lensed quasar, we started a
project to search for large separation lenses in the quasar sample of
the Sloan Digital Sky Survey \citep[SDSS;][]{york00}.  This project
complements ongoing searches for small separation lenses in SDSS
\citep[e.g.,][]{pindor03,inada03a}.  The SDSS has completed less than
half of its planned observations, but already it contains more than
30000 quasars and is superior to previous large separation lens surveys
in several ways.  The full SDSS sample will comprise $\sim$100000
quasars, so we ultimately expect to find several large separation lensed
quasars \citep{keeton01c,takahashi01,li02,kuhlen04}.  One of the most
important advantages of the SDSS in searching for large separation
lensed quasars is that imaging in five broad optical bands allows us
to select lens candidates quite efficiently.

Recently we reported the discovery of the large separation lensed
quasar SDSS~J1004+4112 at $z=1.73$ \citep[][]{inada03b} in the
SDSS.  The quasar itself turned out to be previously identified in the
{\it ROSAT} All Sky Survey \citep*{cao99} and the Two-Micron All-Sky
Survey \citep{barkhouse01}, but was not recognized as a lensed system.
\citet{inada03b} showed that SDSS~J1004+4112 consists of four quasar
images with the same redshift from the Keck spectroscopy.  The colors
of galaxies found by Subaru imaging follow-up observations indicated
the presence of a cluster of galaxies at $z=0.68$.  Moreover, the
configuration of the four images was successfully reproduced by a
simple lens model based on a singular isothermal ellipsoid mass
distribution.  All these results strongly implied that SDSS~J1004+4112
is the first quasar lens system due to a massive cluster-scale object.
In this paper, we describe photometric and spectroscopic follow-up
observations of SDSS~J1004+4112 in detail.  We discuss the spectra
of lensed quasar components, including puzzling differences between
emission lines seen in the different images.  We analyze deep multicolor
imaging data to show the existence of a lensing cluster more robustly.
We also present detailed mass modeling of the lens, and discuss the
implications of this system for the statistics of large separation
lenses.

The paper is organized as follows. Section \ref{sec:select} describes
the method used to identify large separation lens candidates in the SDSS
data. The results of follow-up observations are summarized in
\S\ref{sec:data}. In \S\ref{sec:model} we perform mass modeling of
SDSS~J1004+4112, and in \S\ref{sec:stat} we consider the statistical
implications of the discovery of SDSS~J1004+4112. We summarize our
results and conclusions in \S\ref{sec:discussion}.
Throughout this paper, we assume the popular ``concordance'' cosmology
with $\Omega_M=0.27$, $\Omega_\Lambda=0.73$, and $H_0=70$~km~s$^{-1}$
Mpc$^{-1}$ \citep[e.g.,][]{spergel03}.

%%%%%%%%%%%%%%%%%%%%%%%%%%%%%%%%%%%%%%%%%%%%%%%%%%%%%%%%%%%
%%%%%%%%%%%%%%%%%%%%%%%%%%%%%%%%%%%%%%%%%%%%%%%%%%%%%%%%%%%
\section{Candidate Selection from the SDSS Object Catalog\label{sec:select}}
%%%%%%%%%%%%%%%%%%%%%%%%%%%%%%%%%%%%%%%%%%%%%%%%%%%%%%%%%%%
%%%%%%%%%%%%%%%%%%%%%%%%%%%%%%%%%%%%%%%%%%%%%%%%%%%%%%%%%%%

The SDSS is a survey to
image a quarter of the Celestial Sphere at high Galactic latitude
and to measure spectra of galaxies and quasars found in the
imaging data \citep{blanton03}. The dedicated 2.5-meter telescope
at Apache Point Observatory (APO) is equipped with a multi-CCD
camera \citep{gunn98} with five broad bands centered at $3561$,
$4676$, $6176$, $7494$, and $8873${\,\AA} \citep{fukugita96}. The
imaging data are automatically reduced by a photometric pipeline
\citep{lupton01}. The astrometric positions are accurate to about
$0\farcs1$ for sources brighter than $r=20.5$ \citep{pier03}. The
photometric errors are typically less than 0.03 magnitude
\citep{hogg01,smith02}. The SDSS quasar selection algorithm is presented
in \citet{richards02}. The SDSS spectrographs are used to obtain spectra,
covering $3800$--$9200${\,\AA} at a resolution of $1800$--$2100$, for
the quasar candidates. The public data releases of the SDSS are
described in \citet{stoughton02} and \citet{abazajian03}.

Large separation lens candidates can be identified from the SDSS data
as follows. First, we select objects that were initially identified as
quasars by the spectroscopic pipeline. Specifically, among
SDSS spectroscopic targets we select all objects that have spectral
classification of {\tt SPEC\_QSO} or {\tt SPEC\_HIZ\_QSO} with
confidence {\tt z\_conf} larger than $0.9$ \citep[see][for details of
the SDSS spectral codes]{stoughton02}. Next we check the
colors of nearby unresolved sources to see if any of those sources
could be an additional quasar image, restricting the lens search to
separations $\theta<60''$.  We define a large separation lens by
$\theta>7''$ so that it exceeds the largest image separation lenses
found so far: Q0957+561 with $\theta=6\farcs26$ \citep{walsh79} and
RX~J0921+4529 with $\theta=6\farcs97$ \citep{munoz01}, both of which
are produced by galaxies in small clusters. We regard the stellar object
as a candidate companion image if the following color conditions are
satisfied:
%%%%%%%%%%%%%%%%%%%%%%%%%%%%%%%%%%%%%%%%%%%%%%%%%%%%%%%%%%%%%%%%%%%%%%%%
\begin{eqnarray}
 \left|\Delta(j-k)\right|&<&3\sigma_{\Delta(j-k)}\nonumber\\
&=&3\sqrt{\left(\sigma_{j,{\rm err}}^2+\sigma_{k,{\rm err}}^2\right)_{\rm quasar}+\left(\sigma_{j,{\rm err}}^2+\sigma_{k,{\rm err}}^2\right)_{\rm stellar}},
\end{eqnarray}
%%%%%%%%%%%%%%%%%%%%%%%%%%%%%%%%%%%%%%%%%%%%%%%%%%%%%%%%%%%%%%%%%%%%%%%%
%%%%%%%%%%%%%%%%%%%%%%%%%%%%%%%%%%%%%%%%%%%%%%%%%%%%%%%%%%%%%%%%%%%%%%%%
\begin{equation}
\left|\Delta(j-k)\right|<0.1,
\end{equation}
%%%%%%%%%%%%%%%%%%%%%%%%%%%%%%%%%%%%%%%%%%%%%%%%%%%%%%%%%%%%%%%%%%%%%%%%
%%%%%%%%%%%%%%%%%%%%%%%%%%%%%%%%%%%%%%%%%%%%%%%%%%%%%%%%%%%%%%%%%%%%%%%%
\begin{equation}
\left|\Delta i^*\right|<2.5,
\end{equation}
%%%%%%%%%%%%%%%%%%%%%%%%%%%%%%%%%%%%%%%%%%%%%%%%%%%%%%%%%%%%%%%%%%%%%%%%
where $\{j,k\}=\{u^*,g^*\}$, $\{g^*,r^*\}$, $\{r^*,i^*\}$, and
$\{i^*,z^*\}$,\footnote{The starred magnitudes ($u^*g^*r^*i^*z^*$)
are used to denote still-preliminary 2.5m-based photometry
\citep[see][]{stoughton02}.}
and $\Delta$ denotes the difference between the spectroscopically
identified quasar and the nearby stellar object.  Note that this
selection criterion is tentative; we still do not know much about
large separation lenses, so selection criteria may evolve as we learn
more.

\vspace{0.5cm}
\centerline{{\vbox{\epsfxsize=7.7cm\epsfbox{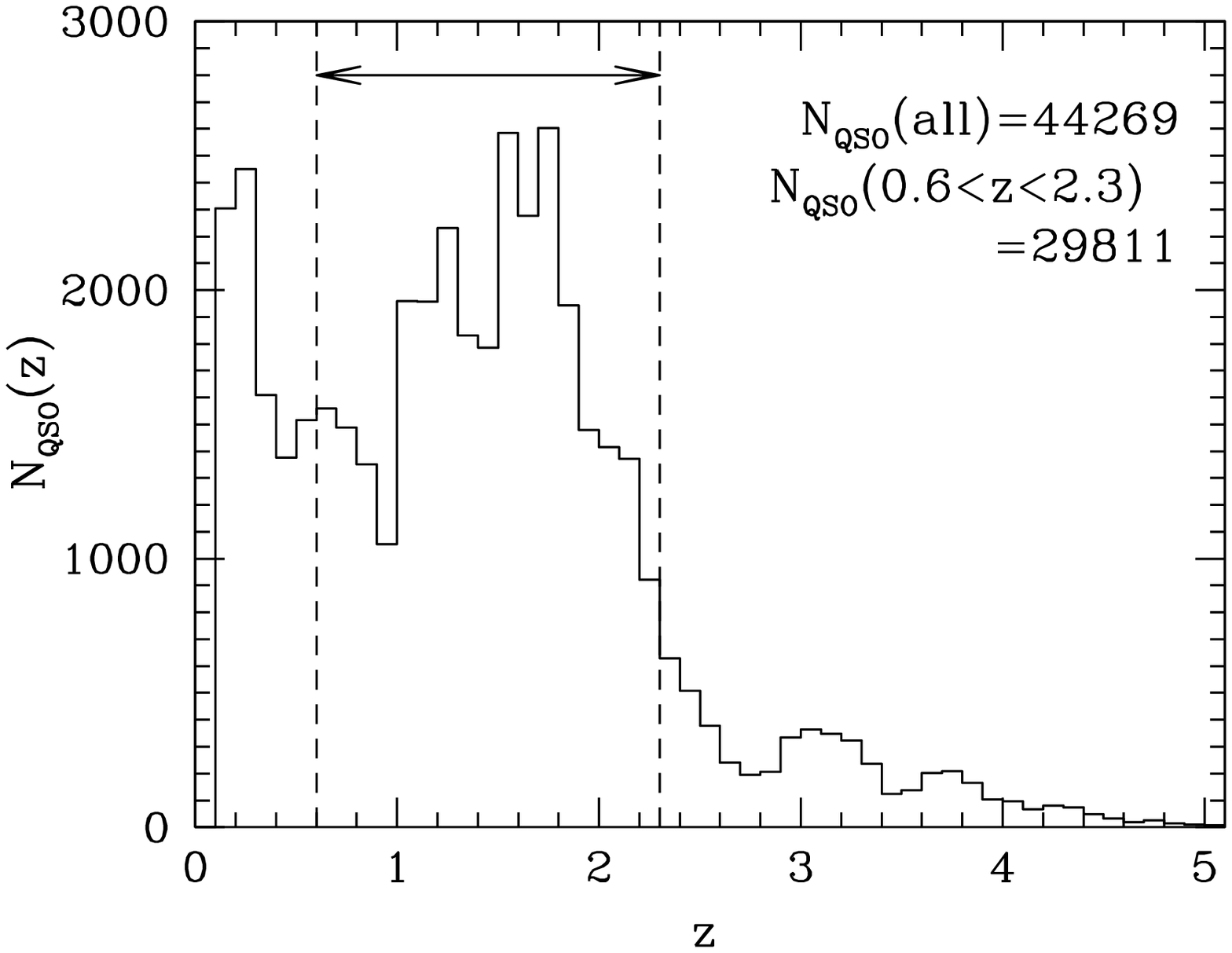}}}}
\figcaption{Redshift distribution of quasars identified by the
 spectroscopic pipeline in the SDSS.  Dashed vertical lines show the
 redshift cut $0.6<z<2.3$ used for the statistical analysis.
 \label{fig:z_qso}}
\vspace{0.5cm}

Our full sample contains 44269 quasars with the redshift distribution
shown in Figure~\ref{fig:z_qso}.  For the lens search we select the
subset of 29811 quasars with $0.6<z<2.3$, making the redshift cuts
for four reasons:
(1) at $z<0.6$ quasars are often extended, which can complicated both
lens searches and also lens statistics analyses;
(2) at $z<0.6$ the sample is contaminated by narrow emission line
galaxies;
(3) at $z>2.3$ we may miss a number of quasar candidates because of
large color errors;
and (4) lens statistics calculations for high redshift quasars are
not very reliable because of uncertainties in the quasar luminosity
function \citep{wyithe02a,wyithe02b}.
Lens surveys of high-redshift quasars are of course very interesting
for insights into the abundance and formation of distant quasars;
a search for lenses among high-redshift SDSS quasars is the subject
of a separate analysis by \citet{richards04}. 

\begin{figure*}[p]
\epsscale{1.2}
\plotone{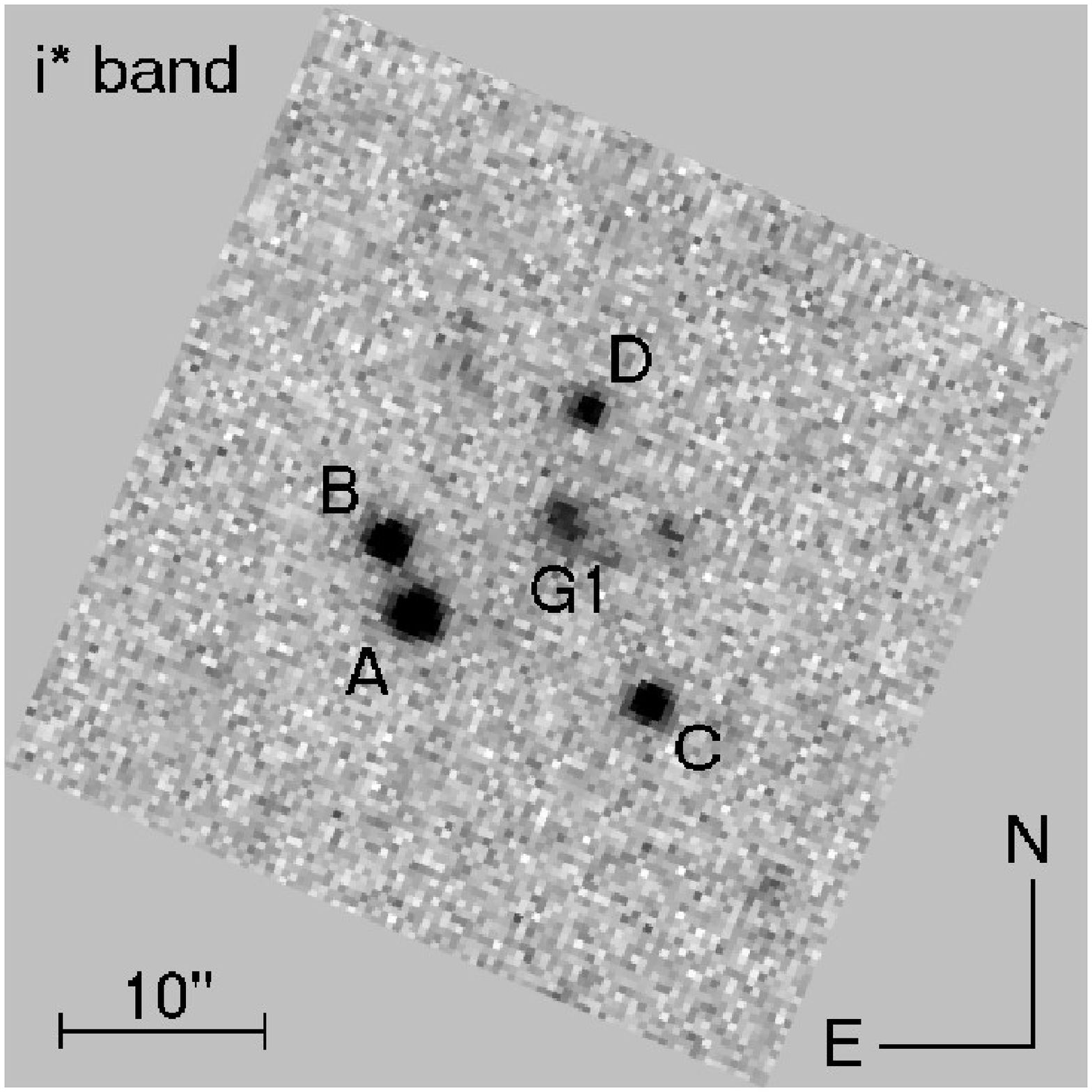}
\caption{SDSS $i^*$-band image of SDSS~J1004+4112. Components A, B,
C, and D are lensed images while component G1 is the brightest galaxy
in the lensing cluster.
 \label{fig:sdss1004_i}}
\end{figure*}

SDSS~J1004+4112 was first selected as a lens candidate based on a
pair of components, A and B (see Figure~\ref{fig:sdss1004_i}), where
B is the SDSS spectroscopic target. Components C and D were
identified by visual inspection and found to have colors similar
to those of A and B (even though they do not match the above color
criteria).  Table~\ref{table:sdssphoto} summarizes the photometry
for the four components, and Table~\ref{table:astrometry} gives the
astrometry for the four components as well as the galaxy G1 seen
in Figure~\ref{fig:sdss1004_i}.  The reason that components C and
D have somewhat different colors from B is still unclear, but it
must be understood in order to discuss the completeness of the lens
survey. The difference might be ascribed to differential absorption
or extinction by intervening material \citep{falco99}, or to
variability in the source on time scales smaller than the time
delays between the images \citep*[e.g.,][]{devries03}, both of
which are effects that become more important as the image
separation grows.

%%%%%%%%%%%%%%%%%%%%%%%%%%%%%%%%%%%%%%%%%%%%%%%%%%%%%%%%%%%
%%%%%%%%%%%%%%%%%%%%%%%%%%%%%%%%%%%%%%%%%%%%%%%%%%%%%%%%%%%
\section{Data Analysis\label{sec:data}}
%%%%%%%%%%%%%%%%%%%%%%%%%%%%%%%%%%%%%%%%%%%%%%%%%%%%%%%%%%%
%%%%%%%%%%%%%%%%%%%%%%%%%%%%%%%%%%%%%%%%%%%%%%%%%%%%%%%%%%%

%%%%%%%%%%%%%%%%%%%%%%%%%%%%%%%%%%%%%%%%%%%%%%%%%%%%%%%%%%%
\subsection{Spectroscopic Follow-up Observations}
%%%%%%%%%%%%%%%%%%%%%%%%%%%%%%%%%%%%%%%%%%%%%%%%%%%%%%%%%%%

%%%%%%%%%%%%%%%%%%%%%%%%%%%%%%%%%%%%%%%%%%%%%%%%%%%%%%%%%%%
\subsubsection{Quasar Images}
%%%%%%%%%%%%%%%%%%%%%%%%%%%%%%%%%%%%%%%%%%%%%%%%%%%%%%%%%%%

\begin{figure*}[p]
\epsscale{1.4}
\plotone{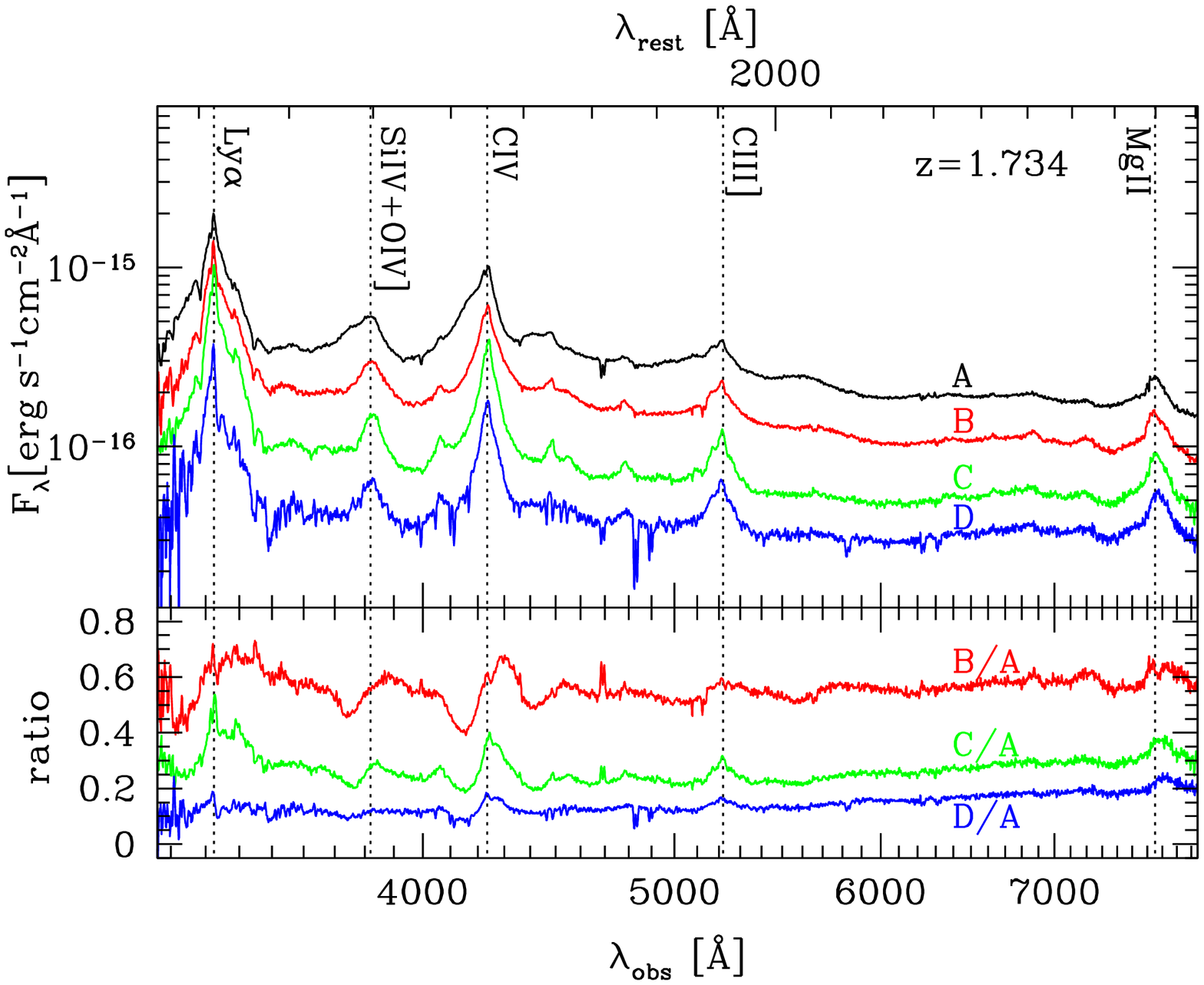}
\caption{Spectra ({\it top}) and flux ratios ({\it bottom}) of
 SDSS~J1004+4112 components A, B, C, and D taken with LRIS on Keck
 I. In the upper panel, we can confirm that all components have
 Ly$\alpha$, \ion{Si}{4}, \ion{C}{4}, \ion{C}{3] }, and \ion{Mg}{2}
 emission lines at $z=1.734$. The flux ratios shown in the lower panel
 are almost constant for a wide range of wavelength. Several absorption
 lines are also seen in the spectra (see text for details).
 \label{fig:spec}}
\end{figure*}

Since only component B has a spectrum from SDSS, we obtained 
spectra of the other components to investigate the lensing hypothesis.
The first spectroscopic follow-up observations were done on 2003 May 2
and 5 with the Double Imaging Spectrograph of the Astrophysical
Research Consortium (ARC) 3.5-m telescope at APO. All four components
have a prominent \ion{C}{4} emission line ($1549.06${\,\AA}) at
$\lambda_{\rm obs} \sim 4230${\,\AA}, indicating that they are
quasars with very similar redshifts. Spectra with higher resolution
and longer wavelength range were taken on 2003 May 30 with the
Low-Resolution Imaging Spectrometer \citep[LRIS;][]{oke95} of the
Keck~I telescope at the W.~M.~Keck Observatory on Mauna Kea, Hawaii,
USA. The blue grism is 400 line mm$^{-1}$, blazed at $3400${\,\AA},
$1.09${\,\AA} pixel$^{-1}$, covering $3000${\,\AA} to $5000${\,\AA}.
The red grating is 300 line mm$^{-1}$, blazed at $5000${\,\AA},
$1.09${\,\AA} pixel$^{-1}$, covering $5000${\,\AA} to the red limit
of the detector. The spectra were obtained with 900 sec exposures
and a $1''$ slit in $0\farcs9$ seeing.  The data were reduced in a
standard way using IRAF.\footnote{IRAF is distributed by the National
Optical Astronomy Observatories, which are operated by the Association
of Universities for Research in Astronomy, Inc., under cooperative
agreement with the National Science Foundation.} The  Keck/LRIS
spectra are shown in Figure~\ref{fig:spec}. All four components show
emission lines of Ly$\alpha$, \ion{Si}{4}, \ion{C}{4}, \ion{C}{3]},
and \ion{Mg}{2}.  They have nearly identical redshifts of $z=1.734$,
with velocity differences less than $50\,{\rm km\,s^{-1}}$
(see Table~\ref{table:sdssphoto}).  The flux ratios between the
images (see Figure~\ref{fig:spec}) are almost constant over the
wavelength range 3000--8000{\,\AA}, indicating that these are actual
lensed images. From the spectra, we conclude that the color differences
found in the SDSS images are mainly caused by differences in the
emission lines (discussed below) and by slightly different continuum
slopes. 

\begin{figure*}[t]
\epsscale{1.4}
\plotone{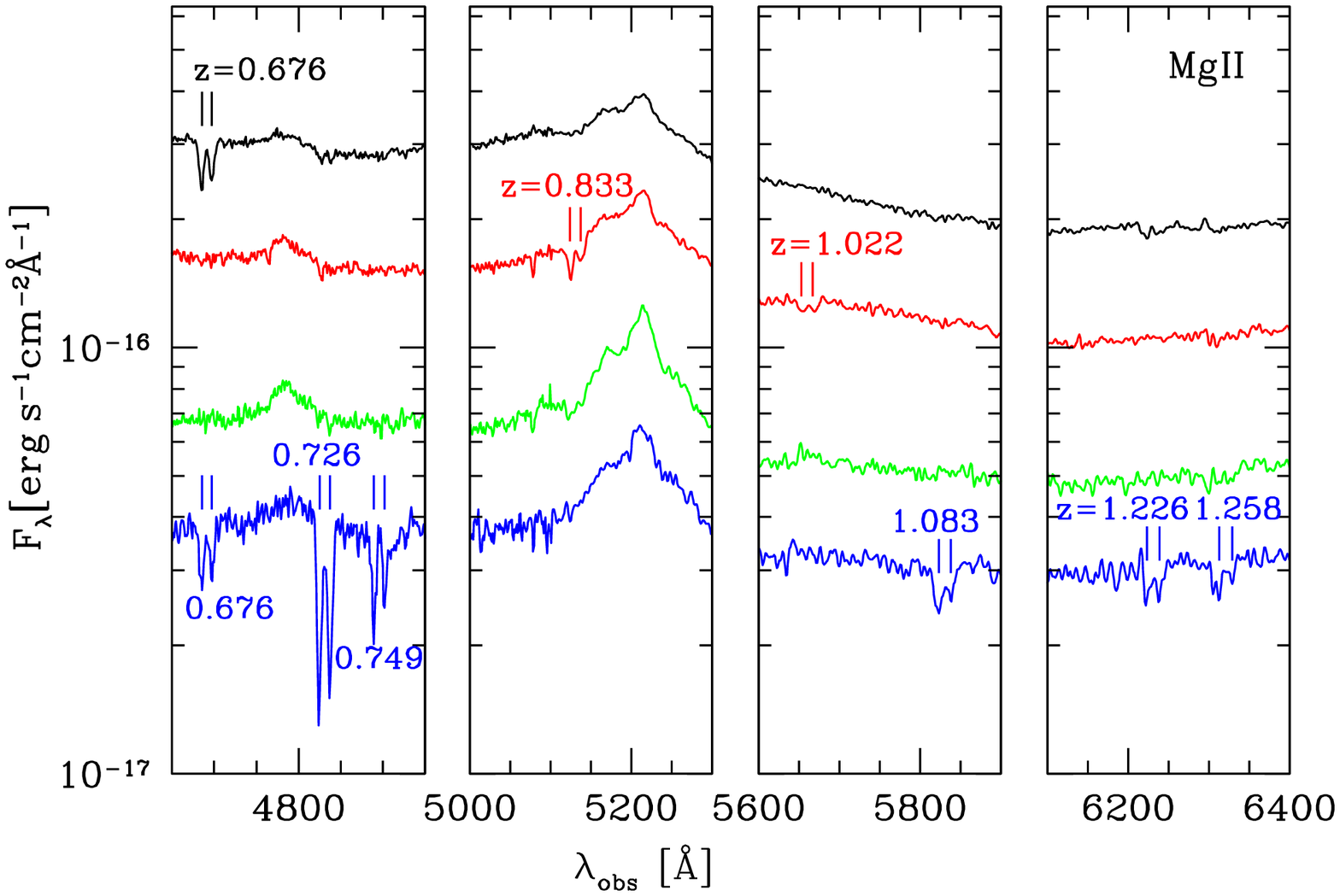}
\caption{The \ion{Mg}{2} doublet absorption lines (rest frame
 wavelengths of $2795.5${\,\AA} and $2802.7${\,\AA}, rest frame
 equivalent width $W_{\rm r}\gtrsim 0.5${\,\AA}) of SDSS~J1004+4112
 components A, B, C, and D at various wavelengths. The absorption lines
 are indicated by vertical lines.
 \label{fig:spec_abs}}
\end{figure*}

Several absorption line systems are seen in the spectra. Components
A and D have intervening \ion{Mg}{2}/\ion{Fe}{2} absorption systems
at $z=0.676$; this redshift is similar to that of the foreground
galaxies (\S\ref{sec:spec_gal}), suggesting that this absorption
system is associated with the lensing galaxies. Component D has
additional \ion{Mg}{2} absorption systems at $z=0.726$, $0.749$,
$1.083$, $1.226$, and $1.258$. Figure~\ref{fig:spec_abs} identifies
the various \ion{Mg}{2} absorbers. We also note that all four
components have \ion{C}{4} absorption lines just blueward of
\ion{C}{4} emission lines (see Figure~\ref{fig:spec_line}). The
velocity difference between the emission and absorption lines is
$\sim\!500\,{\rm km\,s^{-1}}$, so the absorption system is likely to
be associated with the quasar.  The fact that all four components
have \ion{C}{4} absorption lines offers further evidence that
SDSS~J1004+4112 is indeed a gravitational lens. 

\vspace{0.5cm}
\centerline{{\vbox{\epsfxsize=7.7cm\epsfbox{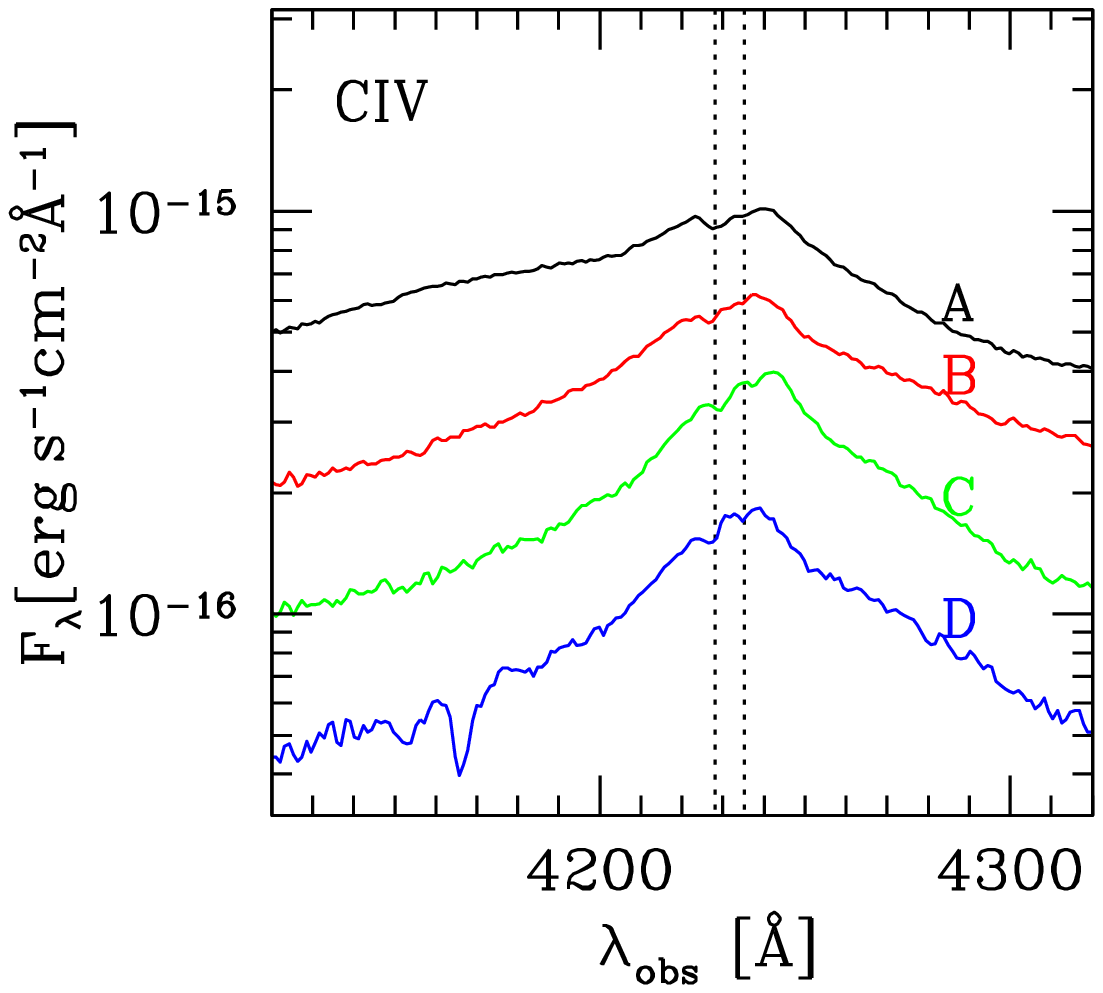}}}}
\figcaption{The \ion{C}{4} lines of SDSS~J1004+4112 components A, B, C,
 and D taken with LRIS. The associated \ion{C}{4} doublet absorption
 lines (rest frame wavelengths $1548.2${\,\AA} and $1550.8${\,\AA},
 denoted by dotted lines) are seen in all four components.
 \label{fig:spec_line}}
\vspace{0.5cm}

Figure~\ref{fig:spec_line} shows notable differences in the \ion{C}{4}
emission line profiles in the different components.  One possible
explanation is the time delay between the lensed images; at any
given observed epoch, the images represent different epochs in the
source frame. However, the fact that the \ion{C}{4} emission lines
in components A and B differ seems to rule out the time delay
explanation: the expected delay (see \S\ref{sec:modelresult}) is
shorter than the month or year time scale on which \ion{C}{4} emission
lines typically vary \citep[e.g.,][]{vandenberk04}. Other possible
explanations include differences between the viewing angles probed
by the images, microlensing amplification of part of the quasar emission
region, significant errors in the predicted time delay between A and B,
or just that the quasar is extremely unusual.  Understanding the
puzzling line profile differences will require further observations,
preferably including measurement of the time delays and spectroscopic
monitoring to identify any variability in the \ion{C}{4} lines.

%%%%%%%%%%%%%%%%%%%%%%%%%%%%%%%%%%%%%%%%%%%%%%%%%%%%%%%%%%%
\subsubsection{Galaxies\label{sec:spec_gal}}
%%%%%%%%%%%%%%%%%%%%%%%%%%%%%%%%%%%%%%%%%%%%%%%%%%%%%%%%%%%

\vspace{0.5cm}
\centerline{{\vbox{\epsfxsize=7.7cm\epsfbox{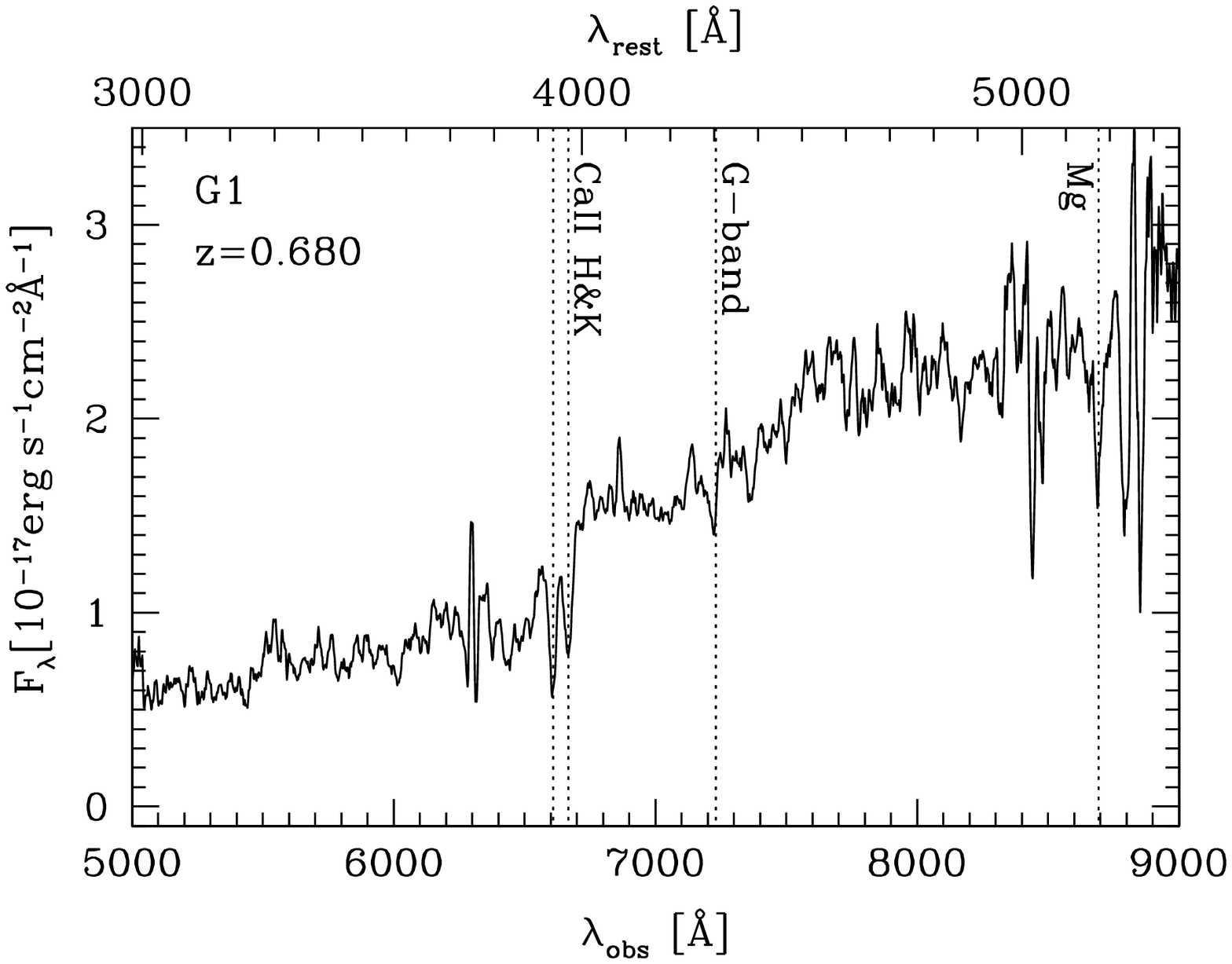}}}}
\figcaption{Spectrum of the galaxy G1 taken with LRIS on Keck
 I. The break, \ion{Ca}{2} H\&K absorption lines, and Mg absorption line
 are consistent with redshift $z=0.680$ ($z=0.6799\pm0.0001$ from
 the \ion{Ca}{2} H line). The G-band also appears in the spectrum. 
 \label{fig:spec_g}}
\vspace{0.5cm}

The spectrum of the galaxy G1, the brightest object near the center of 
the quasar configuration (see Figure~\ref{fig:sdss1004_i}), was
acquired on 2003 May 30 with LRIS.  The spectrum measured from a
900 sec exposure is shown in Figure~\ref{fig:spec_g}. We confirm the
break and \ion{Ca}{2} H\&K lines at $\lambda_{\rm obs}\sim6700${\,\AA}.
The G-band also appears in the spectrum. From the \ion{Ca}{2} H\&K
and Mg lines we derive the redshift of G1 as $z=0.680$. 

\vspace{0.5cm}
\centerline{{\vbox{\epsfxsize=7.7cm\epsfbox{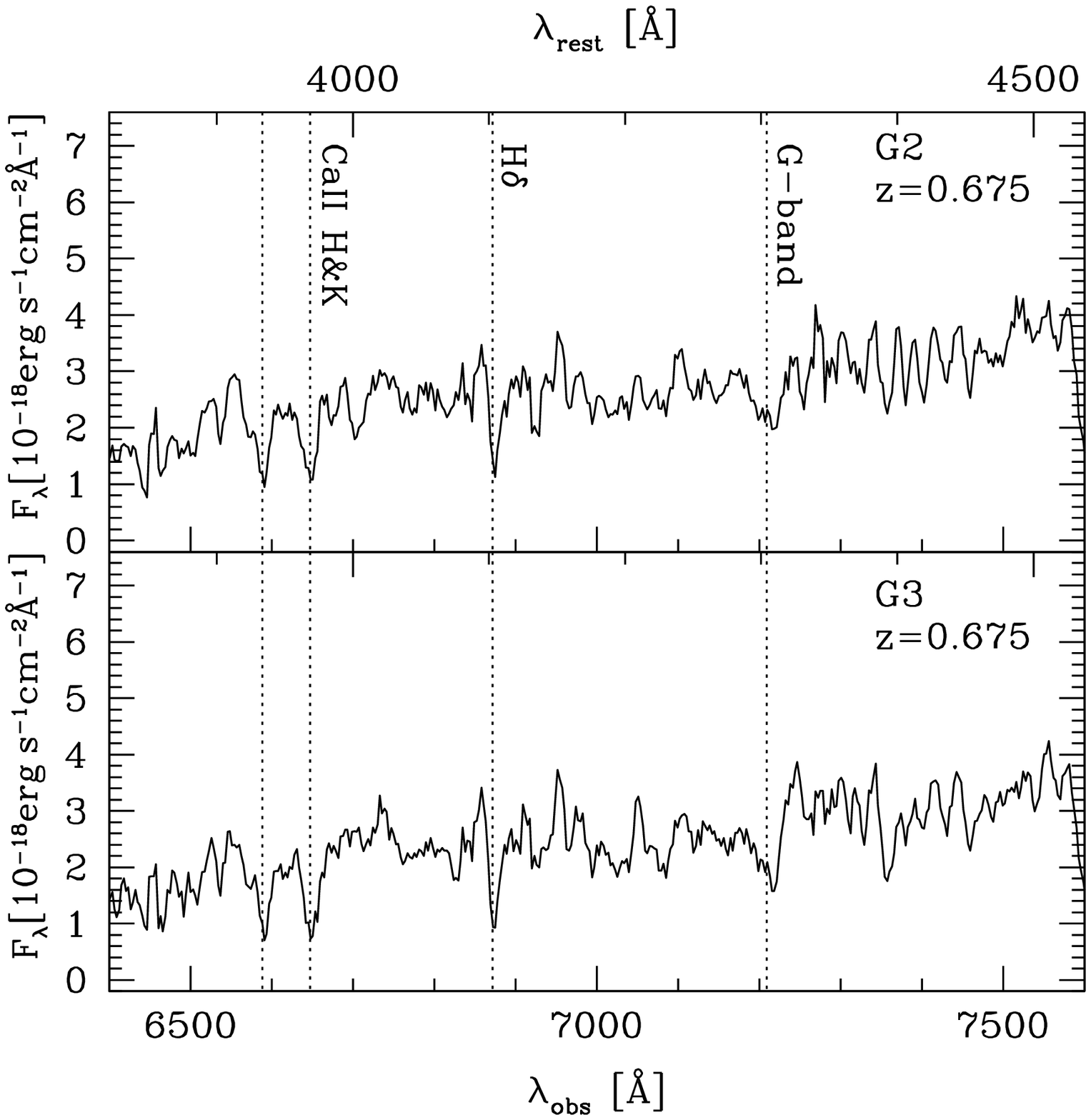}}}}
\figcaption{Spectra of galaxies G2 and G3 taken with FOCAS on the Subaru
 8.2-m telescope. From the absorption lines \ion{Ca}{2} H\&K, H$\delta$,
 and G-band, we find that the redshifts of both galaxies are $z=0.675$
 ($z=0.6751\pm0.0001$ from the H$\delta$ lines).
 \label{fig:spec_focas}}
\vspace{0.5cm}

The spectra of two additional galaxies near G1 (see \S\ref{sec:imaging})
were taken on 2003 June 20 with the Faint Object Camera and Spectrograph
\citep[FOCAS;][]{kashikawa02} on the Subaru 8.2-m telescope of the
National Astronomical Observatory of Japan on Mauna Kea, Hawaii, USA.
We used the 300B grism together with the SY47 filter, and took optical
spectra covering $4100${\,\AA} to $10000${\,\AA} with resolution
$2.84${\,\AA} pixel$^{-1}$. The seeing was $0\farcs7$, and the exposure
time was 1740 sec for both galaxies. The data were reduced in a standard
way using IRAF. The spectra are shown in Figure~\ref{fig:spec_focas}. 
Both galaxies, denoted as G2 and G3, have $z = 0.675$, only
$\sim\!700\,{\rm km\,s^{-1}}$ from the redshift of G1. 

%%%%%%%%%%%%%%%%%%%%%%%%%%%%%%%%%%%%%%%%%%%%%%%%%%%%%%%%%%%
\subsection{Imaging Follow-up Observations\label{sec:imaging}}
%%%%%%%%%%%%%%%%%%%%%%%%%%%%%%%%%%%%%%%%%%%%%%%%%%%%%%%%%%%

\begin{figure*}[p]
\epsscale{1.25}
\plotone{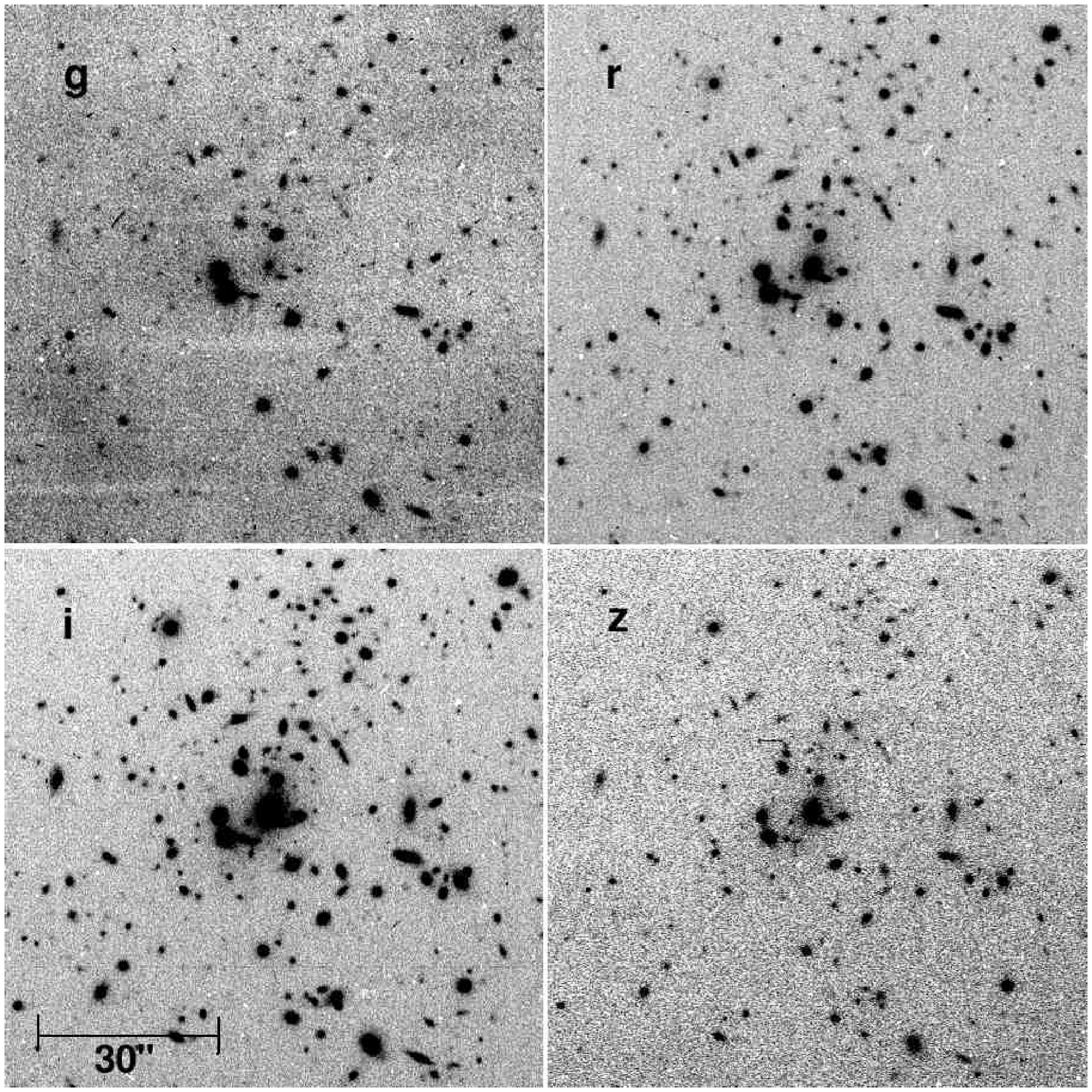}
\caption{Deep $griz$ images taken with Suprime-Cam on the Subaru 8.2-m
 telescope. The exposure details are summarized in Table~\ref{table:subaruobs}.
 \label{fig:subaru_griz}}
\end{figure*}

\begin{figure*}[p]
\epsscale{1.2}
\plotone{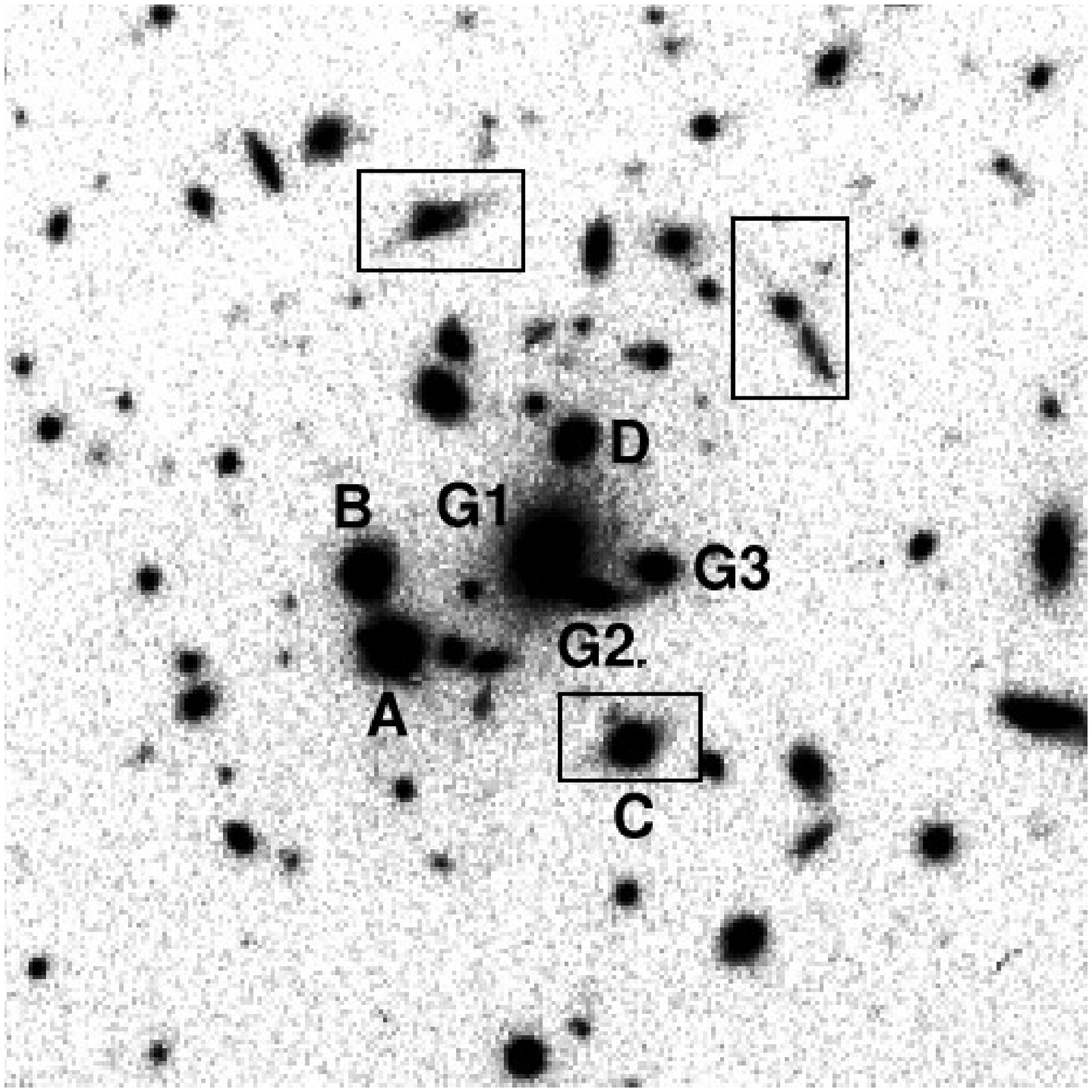}
\caption{The central region of the Suprime-Cam $i$-band image.
 The galaxies with measured redshifts (G1 from LRIS and G2 and G3 from
 FOCAS) as well as the four lensed images are labeled. The possible
 lensed arclets are marked with rectangles.
 \label{fig:subaru_center}}
\end{figure*}

%%%%%%%%%%%%%%%%%%%%%%%%%%%%%%%%%%%%%%%%%%%%%%%%%%%%%%%%%%%
\subsubsection{Observations}
%%%%%%%%%%%%%%%%%%%%%%%%%%%%%%%%%%%%%%%%%%%%%%%%%%%%%%%%%%%

A deep $r$-band image of SDSS~J1004+4112 was taken on 2003 May 5
with the Seaver Prototype Imaging camera of the ARC 3.5-m telescope
at APO. The image shows rich structure, with many galaxies between
and around the quasar components suggesting a possible galaxy cluster
in the field. For a further check, we obtained deeper multi-color
($griz$) images on 2003 May 28 with the Subaru Prime Focus Camera
\citep[Suprime-Cam;][]{miyazaki02} on the Subaru 8.2-m telescope.
The exposure times and limiting magnitudes of the Suprime-Cam images
are given in Table~\ref{table:subaruobs}. Suprime-Cam has a pixel
scale of $0\farcs2\,{\rm pixel^{-1}}$, and the seeing was
$0\farcs5$--$0\farcs6$.  The frames were reduced (bias-subtracted
and flat-field corrected) in a standard way. The resulting images
are shown in Figure~\ref{fig:subaru_griz}. It is clear that there
are many red galaxies around the four images. Moreover, we find
three possible lensed arclets (distorted images of galaxies behind
the cluster), which are shown in more detail in
Figure~\ref{fig:subaru_center}. The fact that the arclets are
relatively blue compared with the brighter galaxies in the field
\citep[see Figure~2 in][]{inada03b} suggests that the arclets may
be images of distant galaxies \citep*[e.g.,][]{colley96}. Confirming
that they are lensed images will require higher resolution images
and measurements of the arclets' redshifts. If the hypothesis is
confirmed, the arclets will provide important additional constraints
on the lens mass distribution.

%%%%%%%%%%%%%%%%%%%%%%%%%%%%%%%%%%%%%%%%%%%%%%%%%%%%%%%%%%%
\subsubsection{Colors of Nearby Galaxies}
%%%%%%%%%%%%%%%%%%%%%%%%%%%%%%%%%%%%%%%%%%%%%%%%%%%%%%%%%%%

\begin{figure*}[tb]
\epsscale{1.23}
\plotone{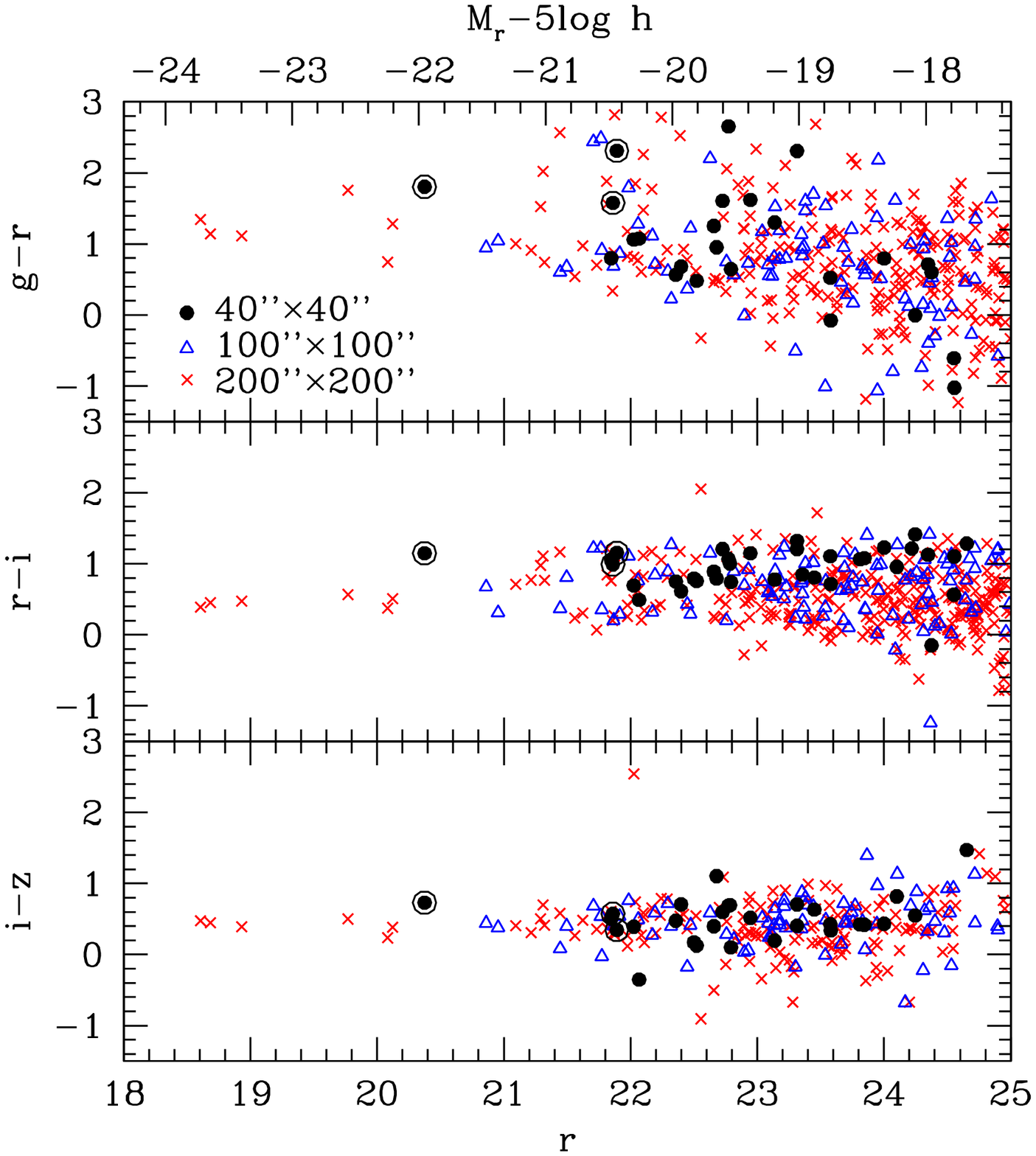}
\caption{Color-magnitude diagrams for the SDSS~J1004+4112 field taken
 with Suprime-Cam. We divide the galaxies into three categories
 according to their positions: filled circles denote galaxies inside a
 $40''\times40''$ box centered on G1, open triangles denote galaxies
 inside a $100''\times100''$ box, and crosses denote galaxies inside a
 $200''\times200''$ box. These box sizes correspond to
 $0.2h^{-1}\,{\rm Mpc}\times0.2h^{-1}{\rm Mpc}$,
 $0.5h^{-1}{\rm Mpc}\times0.5h^{-1}{\rm Mpc}$, and
 $1.0h^{-1}{\rm Mpc}\times1.0h^{-1}{\rm Mpc}$ at $z=0.68$, respectively. 
 Three spectroscopically confirmed member galaxies are marked with
 open circles. The corresponding $r$-band absolute magnitudes at
 $z=0.68$ (without K-correction) are given at the top of the frame.
\label{fig:colormag}}
\end{figure*}

The colors of galaxies in the vicinity of SDSS~J1004+4112 can
help us search for the signature of a cluster.  The central regions
of clusters are dominated by early-type galaxies
\citep[e.g.,][]{dressler80} that show tight correlations among
their photometric properties \citep*{bower92}. These correlations
make it possible to search for clusters using color-magnitude and/or
color-color diagrams \citep{dressler92,gladders00,goto02}.

We measure the colors of galaxies using the deep Suprime-Cam $griz$
images. Object identifications are performed using the Source
Extractor algorithm \citep[SExtractor;][]{bertin96}; we identify
objects with SExtractor parameter {\tt CLASS\_STAR} smaller than
0.6 in the $i$ band image as galaxies. Note that this star/galaxy
separation criterion is successful only for objects with
$i \lesssim 24$. The magnitudes in the images are calibrated using
nearby stars whose magnitudes are taken from the SDSS photometric data.

Since the red galaxies in clusters are dominant in the central regions,
and the center of the cluster is thought to be near G1, we divide the
galaxies in the field into three categories: galaxies inside a
$40''\times40''$ (corresponding to $0.2h^{-1}{\rm Mpc}\times0.2h^{-1}{\rm Mpc}$
at $z=0.68$) box centered on G1; galaxies inside a $100''\times100''$
($0.5h^{-1}{\rm Mpc}\times0.5h^{-1}{\rm Mpc}$) box (except for those in
the first category); and galaxies inside a $200''\times200''$
($1.0h^{-1}{\rm Mpc}\times1.0h^{-1}{\rm Mpc}$) box (except for those in
the first two categories).  Figure~\ref{fig:colormag} shows
color-magnitude diagrams for the three categories. It is clear that the
color-magnitude relations, particularly $r-i$ and $i-z$, show tight
correlations for galaxies inside the $40''\times40''$ box. Ridge lines
at $r-i \sim 1.1$ and $i-z \sim 0.5$ strongly suggest a cluster of
galaxies at $z \sim 0.6$ \citep{goto02}. The result is consistent with
the Keck and Subaru spectroscopic results showing that the redshifts
of galaxies G1, G2, and G3 are all $z\sim0.68$.

\begin{figure*}[t]
\epsscale{1.4}
\plotone{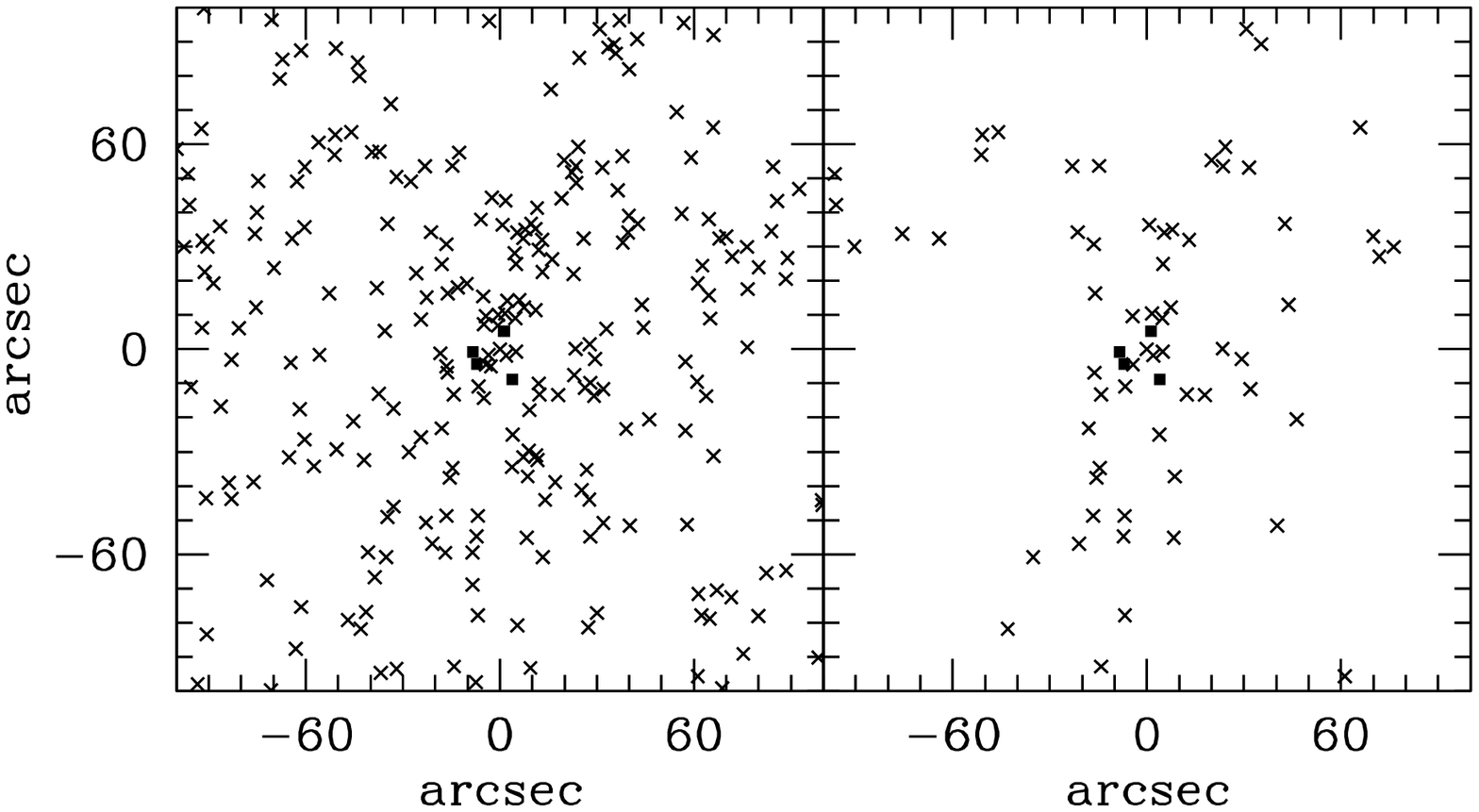}
\caption{Distributions of galaxies brighter than $i=24$ with
 ({\it right}) and without ({\it left}) the color cut. The origin
 $(0,0)$ is set to the position of the central galaxy G1. Filled
 squares denote the four lensed images.
 \label{fig:colorcut}}
\end{figure*}

We identify cluster members by their location in color-color space
\citep{dressler92,goto02}. We show $g-r-i$ and $r-i-z$ color-color
diagrams in Figures \ref{fig:cc_gri} and \ref{fig:cc_riz}, respectively.
We restrict the plots to galaxies brighter than $i=24$ because of the
limitation of the star/galaxy separation.  We make color-color cuts
based on the colors expected of elliptical galaxies \citep*{fukugita95}:
$g-r>1.5$, $r-i>0.7$, and $i-z>0.2$ for elliptical galaxies at
$z \gtrsim 0.5$. The galaxy distributions with and without the color
cuts are shown in Figure~\ref{fig:colorcut}.  The galaxies that
survive the color cuts are concentrated around G1, so we conclude
that they are candidate members of a cluster of galaxies at $z=0.68$
whose center is near G1.  We note that the distribution of candidate
cluster members is not spherical and appears to be elongated North--South.

\vspace{0.5cm}
\centerline{{\vbox{\epsfxsize=7.7cm\epsfbox{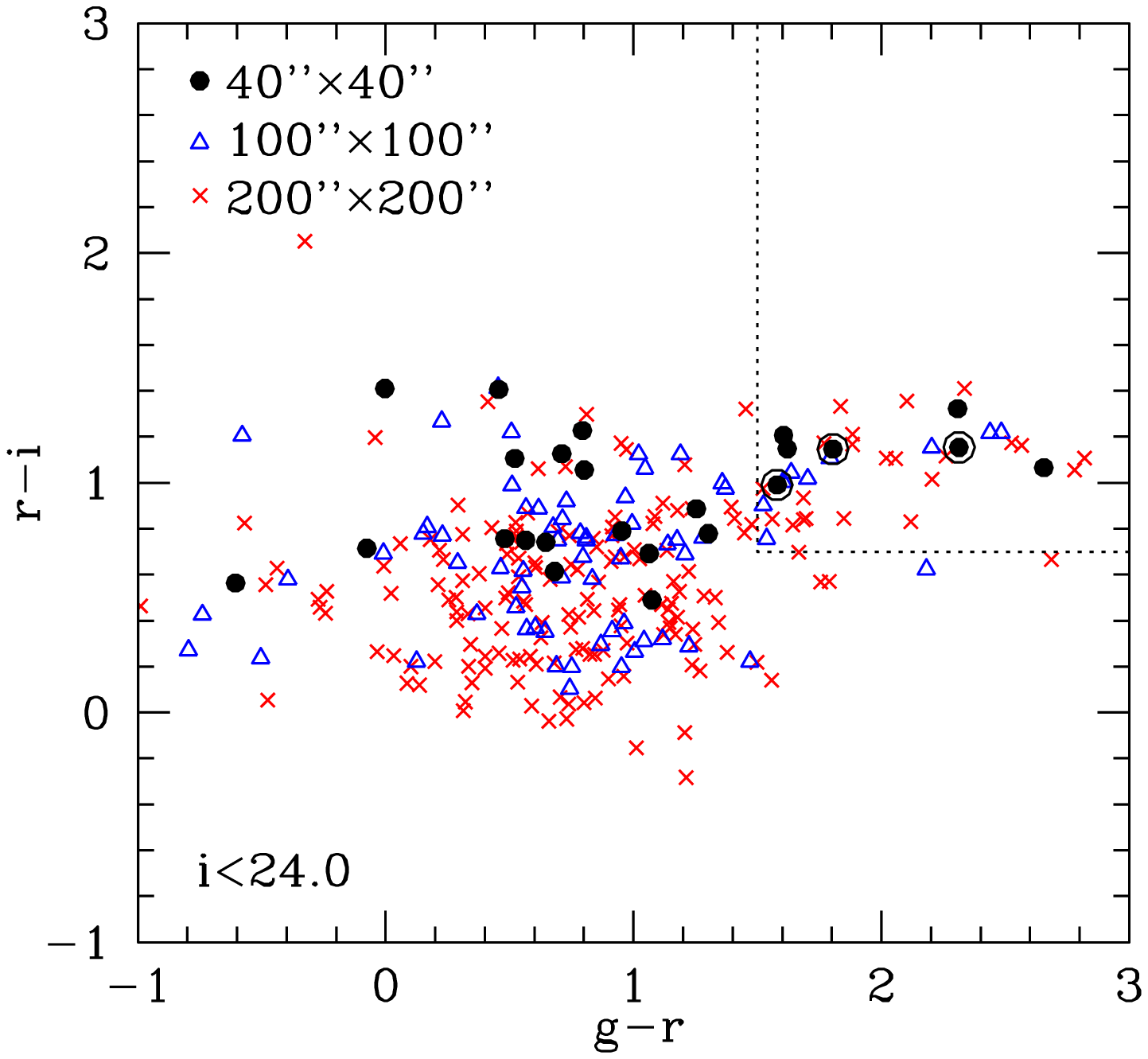}}}}
\figcaption{The $g-r-i$ color-color diagram of galaxies brighter than
 $i=24$. Symbols are the same as in Figure~\ref{fig:colormag}. Dotted
 lines indicate color cuts to find cluster members.
 \label{fig:cc_gri}}
\vspace{0.5cm}

\vspace{0.5cm}
\centerline{{\vbox{\epsfxsize=7.7cm\epsfbox{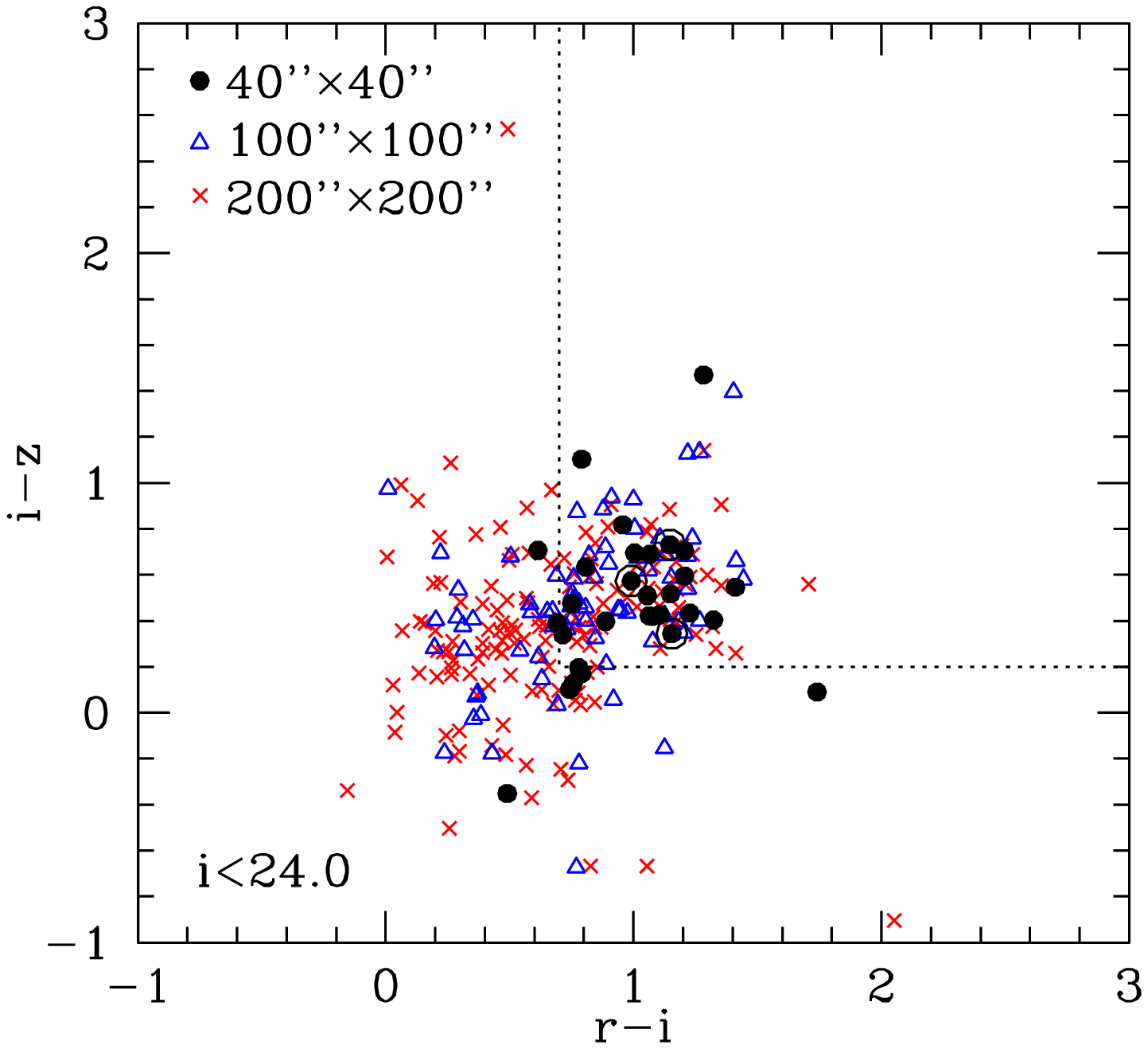}}}}
\figcaption{Similar to Figure~\ref{fig:cc_gri}, but for $r-i-z$.
 \label{fig:cc_riz}}
\vspace{0.5cm}

%%%%%%%%%%%%%%%%%%%%%%%%%%%%%%%%%%%%%%%%%%%%%%%%%%%%%%%%%%%
%%%%%%%%%%%%%%%%%%%%%%%%%%%%%%%%%%%%%%%%%%%%%%%%%%%%%%%%%%%
\section{Mass Modeling\label{sec:model}}
%%%%%%%%%%%%%%%%%%%%%%%%%%%%%%%%%%%%%%%%%%%%%%%%%%%%%%%%%%%
%%%%%%%%%%%%%%%%%%%%%%%%%%%%%%%%%%%%%%%%%%%%%%%%%%%%%%%%%%%

%%%%%%%%%%%%%%%%%%%%%%%%%%%%%%%%%%%%%%%%%%%%%%%%%%%%%%%%%%%
\subsection{One-component Models\label{sec:onemodel}}
%%%%%%%%%%%%%%%%%%%%%%%%%%%%%%%%%%%%%%%%%%%%%%%%%%%%%%%%%%%

To search for mass models that can explain the image configuration
of SDSS~J1004+4112, we use standard lens modeling techniques as
implemented in the software of \citet{keeton01b}.  The main
constraints come from the image positions.  We also use the flux
ratios as constraints, although we broaden the errorbars to 20\%
to account for possible systematic effects due to source variability
and time delays, micro- or milli-lensing, or differential extinction
(See Table~\ref{table:posflux} for the full set of constraint data).
In particular, the different colors of the images and the different
absorption features seen in Figure~\ref{fig:spec} suggest that
differential extinction may be a significant effect.  In this
section we do {\it not\/} use the position of the main galaxy as
a constraint, because we want to understand what constraints can
be placed on the center of the lens potential from the lens data
alone.

We first consider the simplest possible models for a 4-image lens:
an isothermal lens galaxy with a quadrapole produced either
by ellipticity in the galaxy or by an external shear.  A spherical
isothermal lens galaxy has surface mass density
%%%%%%%%%%%%%%%%%%%%%%%%%%%%%%%%%%%%%%%%%%%%%%%%%%%%%%%%%%%%%%%%%%%%%%%%
\begin{equation}
  \kappa(r) = \frac{\Sigma(r)}{\Sigma_{\rm crit}}
  = \frac{r_{\rm ein}}{2r}\ ,
\end{equation}
%%%%%%%%%%%%%%%%%%%%%%%%%%%%%%%%%%%%%%%%%%%%%%%%%%%%%%%%%%%%%%%%%%%%%%%%
where $r_{\rm ein}$ is the Einstein radius of the lens, and
$\Sigma_{\rm crit}=(c^2/4\pi G)(D_{\rm OS}/D_{\rm OL}D_{\rm LS})$
is the critical surface mass density for lensing, with $D_{\rm OL}$,
$D_{\rm OS}$, and $D_{\rm LS}$ being angular diameter distances from
the observer to the lens, from the observer to the source, and from
the lens to the source, respectively.  The Einstein radius is
related to the velocity dispersion $\sigma$ of the galaxy by
%%%%%%%%%%%%%%%%%%%%%%%%%%%%%%%%%%%%%%%%%%%%%%%%%%%%%%%%%%%%%%%%%%%%%%%%
\begin{equation}
  r_{\rm ein} = 4\pi\,\left(\frac{\sigma}{c}\right)^2\,
    \frac{D_{\rm LS}}{D_{\rm OS}}\ .
\end{equation}
%%%%%%%%%%%%%%%%%%%%%%%%%%%%%%%%%%%%%%%%%%%%%%%%%%%%%%%%%%%%%%%%%%%%%%%%
For an elliptical model we replace $r$ with
$r[1+((1-q^2)/(1+q^2))\cos 2(\theta-\theta_e)]^{1/2}$ in the
surface density, where $q$ and $\theta_e$ are the axis ratio
and position angle of the ellipse.

Simple models using either pure ellipticity or pure shear fail
miserably, yielding $\chi^2$ values no better than $2\times10^{4}$
for $N_{\rm dof} = 4$ degrees of freedom.  This failure is not
surprising: most 4-image lenses require {\it both\/} ellipticity
and external shear \citep*[e.g.,][]{keeton97}, and such a situation
is likely in SDSS~J1004+4112 since the main galaxy is observed to
be elongated and the surrounding cluster surely contributes a shear.

\begin{figure*}[t]
\epsscale{1.4}
\plotone{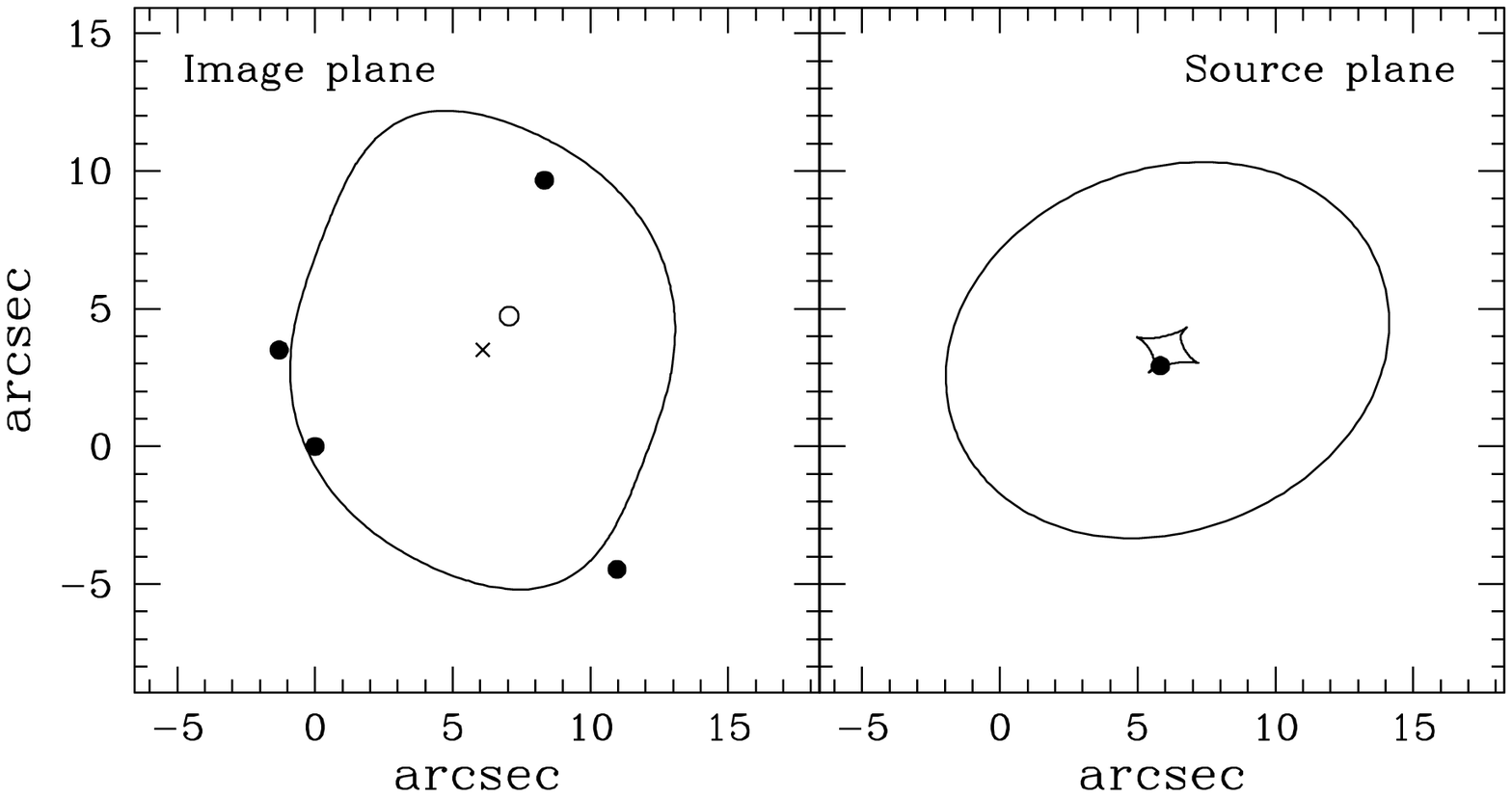}
\caption{Critical curve ({\it left}) and caustic ({\it right}) for the
 best-fit SIE+shear lens model of SDSS~J1004+4112.  In the left panel,
 the filled circles mark the image positions, the open circle indicates
 the observed position of the brightest cluster galaxy G1, and the cross
 marks the best-fit deflector position.  In the right panel the filled
 circle marks the inferred source position.
\label{fig:sieg-crit}}
\end{figure*}

\begin{figure*}[t]
\epsscale{1.4}
\plotone{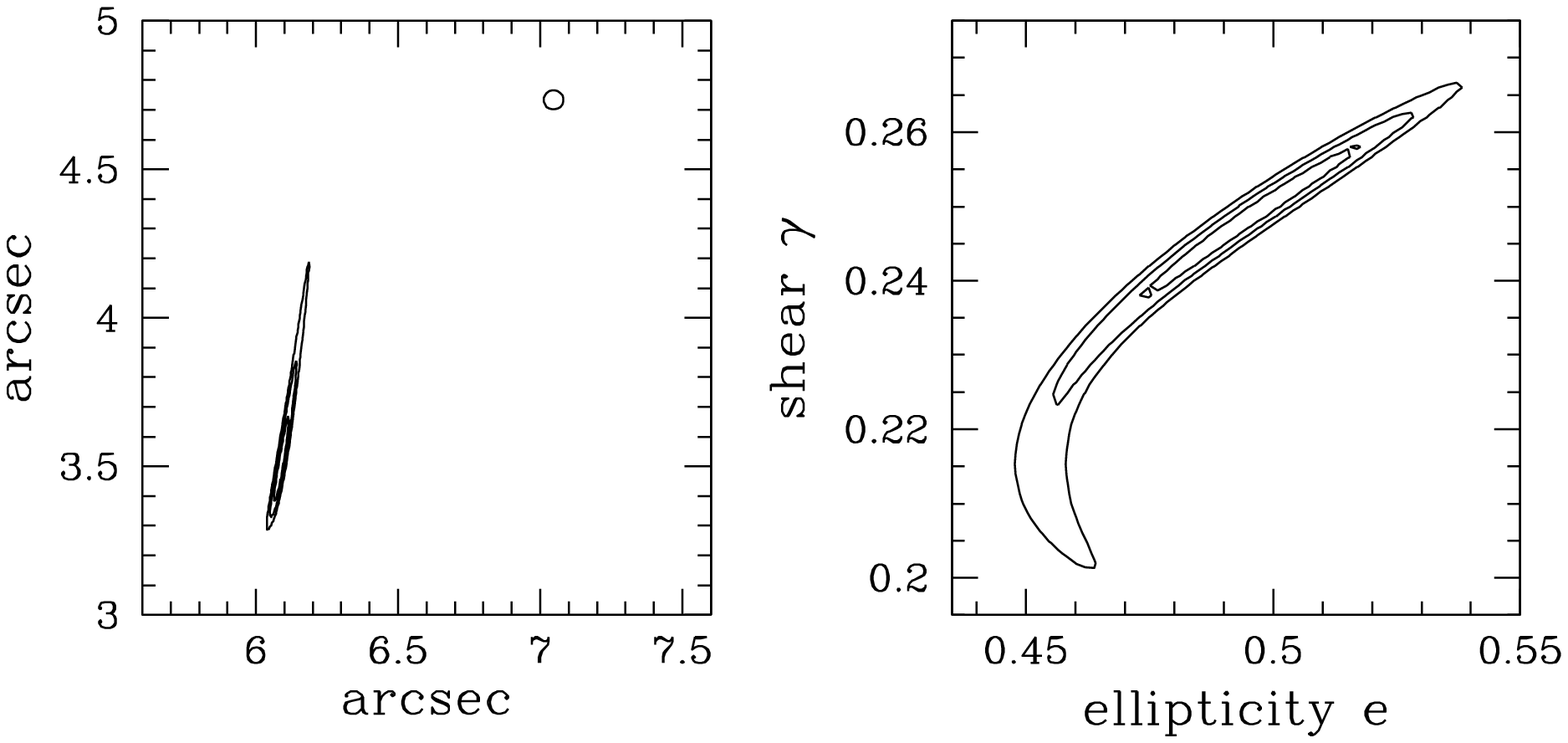}
\caption{Likelihood contours drawn at 1$\sigma$, 2$\sigma$, and
 3$\sigma$ for various parameter combinations in SIE+shear lens models.
 The left panel shows constraints on the position of the deflector;
 the circle marks the observed position of the main galaxy. The right
 panel shows contours in the ellipticity--external shear plane.
\label{fig:sieg-params}}
\end{figure*}

We therefore try models consisting of a singular isothermal ellipsoid
(SIE) plus an external shear $\gamma$.  Even though such models are
still comparatively simple, they can fit the data very well with a
best-fit value of $\chi^2 = 0.33$ for $N_{\rm dof} = 2$.  The best-fit
model has an Einstein radius $r_{\rm ein} = 6\farcs9 = 35\,h^{-1}$~kpc
corresponding to a velocity dispersion of 700~km~s$^{-1}$, an
ellipticity $e=0.50$ at position angle $\theta_e=21\fdg4$, and an
external shear $\gamma=0.25$ at position angle $\theta_\gamma=-60\fdg9$.
Among other known lenses, such a large shear is found only in lenses
lying in cluster environments \citep{burud98,barkana99}.
Figure~\ref{fig:sieg-crit} shows the critical curves and caustics for
the best-fit model.  The inferred source position lies very close to the
caustic and fairly near a cusp, implying that the total magnification is
$\sim$57. Figure~\ref{fig:sieg-params} shows the allowed ranges for the
position of the deflector and the ellipticity and external shear in the
model. The models indicate a small but significant offset of
$1\farcs6 = 7.9\,h^{-1}$~kpc between the center of the lens potential
and the main galaxy, although it remains to be seen whether that
offset is real or an artifact of these still simple lens models.

%%%%%%%%%%%%%%%%%%%%%%%%%%%%%%%%%%%%%%%%%%%%%%%%%%%%%%%%%%%
\subsection{Two-component Models}
%%%%%%%%%%%%%%%%%%%%%%%%%%%%%%%%%%%%%%%%%%%%%%%%%%%%%%%%%%%

Even though the simple SIE+shear model provides a good fit to the
data, we believe that it is not physically plausible because the
system clearly has multiple mass components and it seems unlikely
that all of the mass is associated with a single $\sim$700~km~s$^{-1}$
isothermal component.  The next level of complication is to add a
mass component representing the cluster halo.  We still model the
galaxy G1 explicitly, treating it as an isothermal ellipsoid
constrained by its observed position.  At this point we do not
further complicate the model by attempting to treat the other
galaxies within the lens explicitly.

%%%%%%%%%%%%%%%%%%%%%%%%%%%%%%%%%%%%%%%%%%%%%%%%%%%%%%%%%%%
\subsubsection{Methods}
%%%%%%%%%%%%%%%%%%%%%%%%%%%%%%%%%%%%%%%%%%%%%%%%%%%%%%%%%%%

We model the cluster component with an NFW profile which has been
predicted in cosmological $N$-body simulations \citep*{navarro96,navarro97}:
%%%%%%%%%%%%%%%%%%%%%%%%%%%%%%%%%%%%%%%%%%%%%%%%%%%%%%%%%%%%%%%%%%%%%%%%
\begin{equation}
 \rho(r)=\frac{\rho_{\rm crit}(z)\delta_{\rm c}(z)}
{\left(r/r_{\rm s}\right)\left(1+r/r_{\rm s}\right)^2},
\label{nfw}
\end{equation}
%%%%%%%%%%%%%%%%%%%%%%%%%%%%%%%%%%%%%%%%%%%%%%%%%%%%%%%%%%%%%%%%%%%%%%%%
where $r_{\rm s}$ is a scale radius, $\delta_{\rm c}$ is a characteristic
overdensity (which depends on redshift), and $\rho_{\rm crit}(z)$ is
the critical density of the universe.  Although the NFW density profile
appears to deviate from the results of more recent $N$-body simulations
in the innermost region
\citep*{moore99,ghigna00,jing00a,klypin01,fukushige97,fukushige01,fukushige03a,power03,fukushige04,hayashi04},
we adopt this form for simplicity.  The lensing properties of a spherical
NFW halo are described by the lens potential
\citep*{bartelmann96,golse02,meneghetti03a}
%%%%%%%%%%%%%%%%%%%%%%%%%%%%%%%%%%%%%%%%%%%%%%%%%%%%%%%%%%%%%%%%%%%%%%%%
\begin{equation}
\phi(r) = 2\,\kappa_{\rm s}\,r_{\rm s}^2 \left[
\ln^2\frac{r}{2r_{\rm s}}-\mbox{arctanh}^2\sqrt{1-(r/r_{\rm s})^2} \right],
\end{equation}
%%%%%%%%%%%%%%%%%%%%%%%%%%%%%%%%%%%%%%%%%%%%%%%%%%%%%%%%%%%%%%%%%%%%%%%%
where the lensing strength is specified by the parameter
%%%%%%%%%%%%%%%%%%%%%%%%%%%%%%%%%%%%%%%%%%%%%%%%%%%%%%%%%%%%%%%%%%%%%%%%
\begin{equation}
  \kappa_{\rm s} = \frac{r_{\rm s} \delta_{\rm c}(z) \rho_{\rm crit}(z)}
    {\Sigma_{\rm crit}}.
\label{kappas}
\end{equation}
%%%%%%%%%%%%%%%%%%%%%%%%%%%%%%%%%%%%%%%%%%%%%%%%%%%%%%%%%%%%%%%%%%%%%%%%
Since asphericity in the cluster potential is important in modeling
this system, we generalize the spherical model by adopting elliptical
symmetry in the potential.  Making the potential (rather than the
density) elliptical makes it possible to compute the lensing properties
of an NFW halo analytically \citep{golse02,meneghetti03a}.  We may
still be over-simplifying the mass model, because the cluster profile
may have been modified from the NFW form by baryonic processes such
as gas cooling \citep{rees77,blumenthal86}, and the cluster may have
a complex angular structure if it is not relaxed
\citep[e.g.,][]{meneghetti03a}. To allow for the latter possibility, we
still include a tidal shear in the lens model that can approximate the
effects of complex structure in the outer parts of the cluster.
Overall, our goal is not to model all of the complexities of the lens
potential, but to make the minimal realistic model and see what we can
learn.

\begin{figure*}[t]
\epsscale{1.5}
\plotone{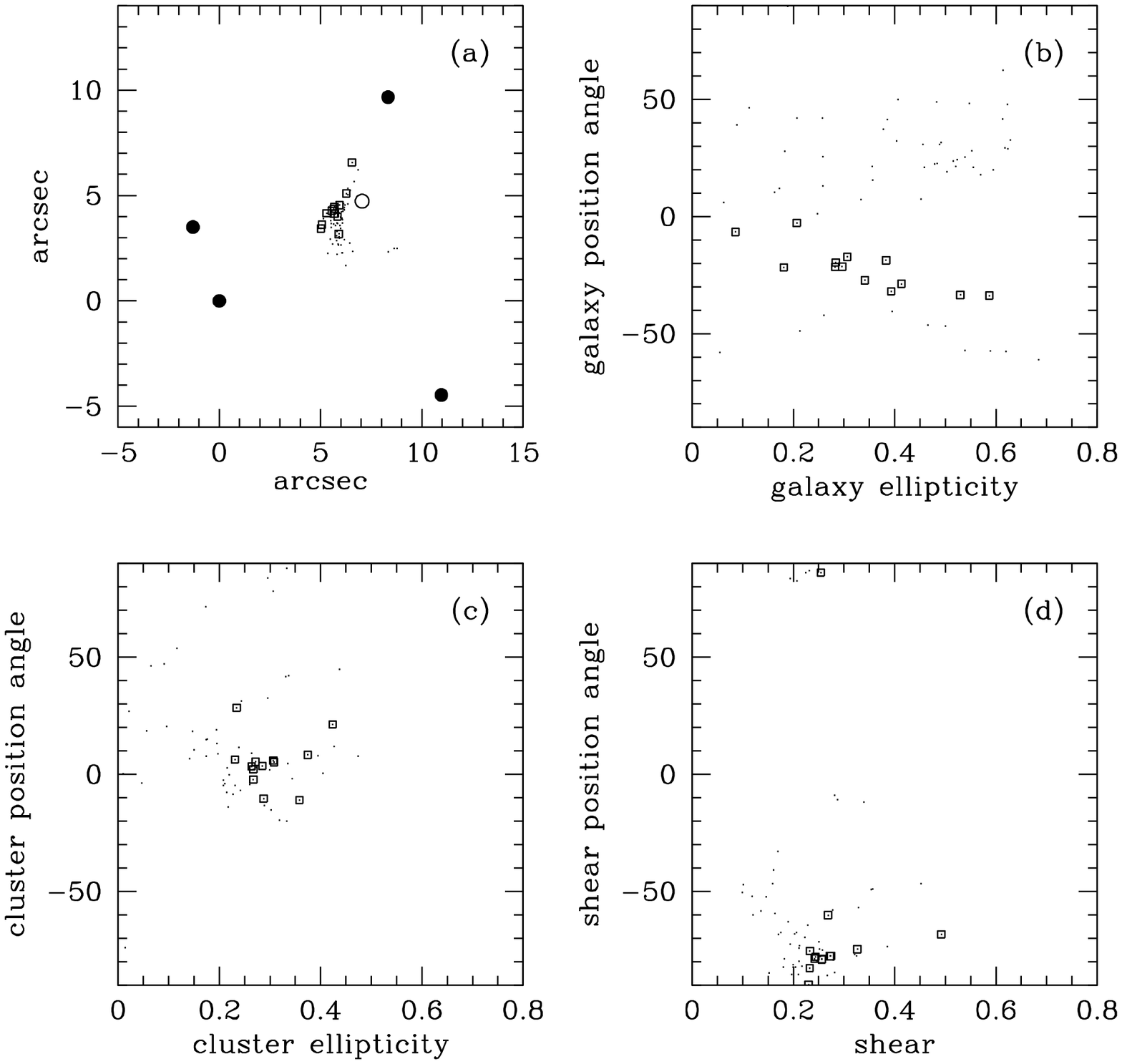}
\caption{Allowed parameter ranges for galaxy+cluster lens models with
a cluster scale radius $r_{\rm s}=40''$. (a) The position of the
cluster component.  The filled circles mark the image positions,
and the open circle marks the observed brightest cluster galaxy G1.
(b) The ellipticity and position angle of the galaxy component.
(c) The ellipticity and position angle of the cluster component.
(d) The amplitude and position angle of the external shear.
Small points show all models, while boxes mark models where the
model galaxy is roughly aligned with the observed galaxy
($\theta_e = -19\fdg9\pm20\fdg0$).
\label{fig:clus-params}}
\end{figure*}

Even with our simplifying assumptions, we still have a complex
parameter space with 11 parameters defining the lens potential:
the mass, ellipticity, and position angle for the galaxy G1;
the position, mass, scale radius, ellipticity, and position angle
for the cluster; and the amplitude and position angle of the shear.
There are also three parameters for the source (position and flux).
With just 12 constraints (position and flux for each of four images),
the models are under-constrained.  We therefore expect that there
may be a range of lens models that can fit the data.  To search
the parameter space and identify the range of models, we follow
the technique introduced by \citet{keeton03} for many-parameter
lens modeling.  Specifically, we pick random starting points in
the parameter space and then run an optimization routine to find
a (local) minimum in the $\chi^2$ surface.  Repeating that process
numerous times should reveal different minima and thereby sample
the full range of models.  Many of the recovered models actually
lie in local minima that do not represent good fits to the data,
so we only keep recovered models with $\chi^2 < 11.8$ \citep[which
represents the 3$\sigma$ limit relative to a perfect fit when
examining two-dimensional slices of the allowed parameter
range; see][]{press92}.

We make one further cut on the models.  From the previous section,
we know that an SIE+shear lens model can give a good fit to the
data.  Thus, there are acceptable two-component models where most
or all of the mass is in the galaxy component and the cluster
contribution is negligible.  To exclude such models as physically
implausible, we impose an upper limit on the velocity dispersion
of the model galaxy.  Specifically, we only keep models with
$\sigma_{\rm gal} < 400$~km~s$^{-1}$, because there are essentially
no galaxies in the observed universe with larger velocity
dispersions, even in rich clusters
\citep[e.g.,][]{kelson02,bernardi03,sheth03}.  Formally, we impose
this cut as an upper limit $r_{\rm ein} < 2\farcs25$ on the
Einstein radius of the galaxy G1.

%%%%%%%%%%%%%%%%%%%%%%%%%%%%%%%%%%%%%%%%%%%%%%%%%%%%%%%%%%%
\subsubsection{Results\label{sec:modelresult}}
%%%%%%%%%%%%%%%%%%%%%%%%%%%%%%%%%%%%%%%%%%%%%%%%%%%%%%%%%%%

\begin{figure*}[t]
\epsscale{1.6}
\plotone{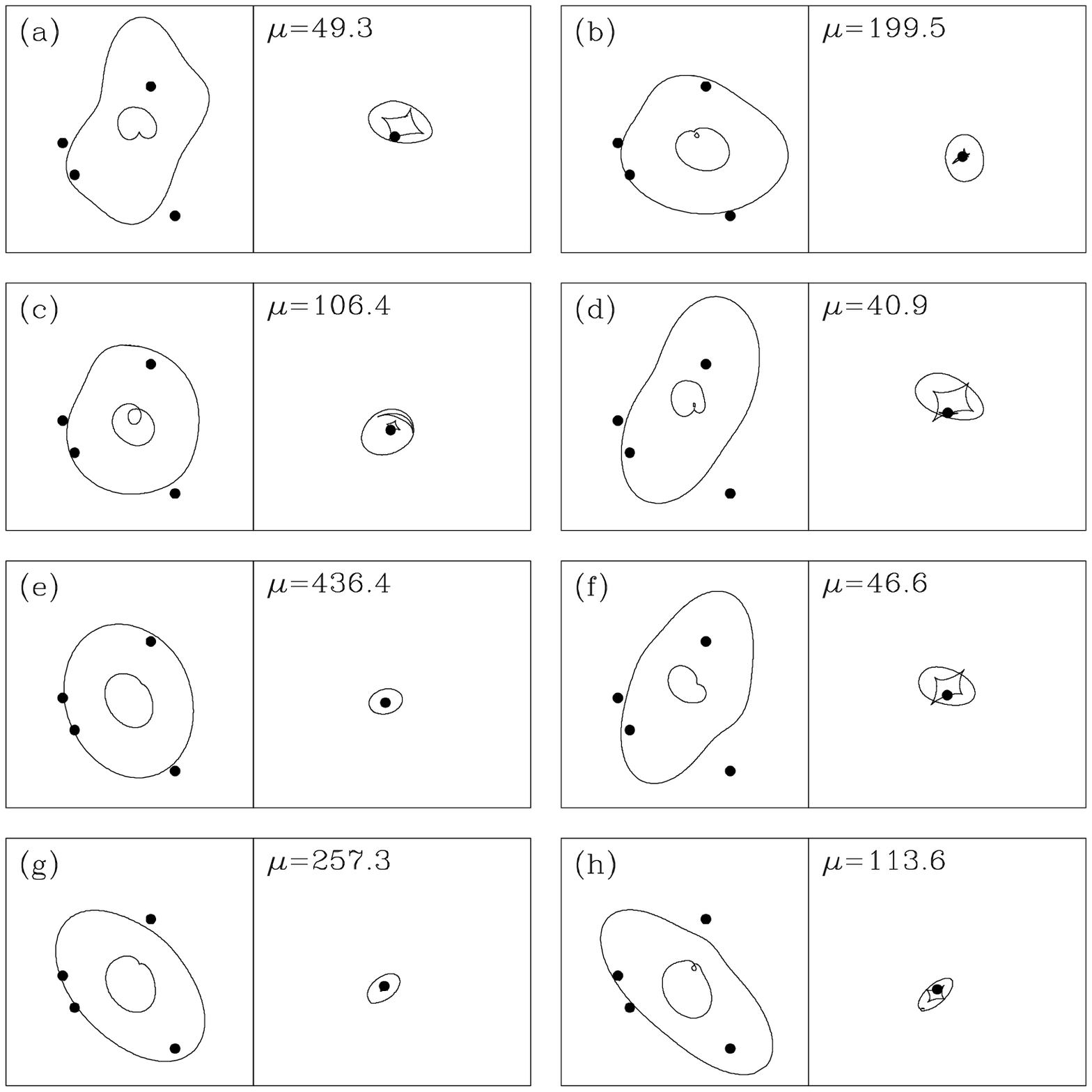}
\caption{Critical curves and caustics for sample galaxy+cluster lens
 models with a cluster scale radius $r_{\rm s}=40\arcsec$.  In each
 panel, the left-hand side shows the critical curves in the image plane,
 and the right-hand side shows the caustics in the source plane on the
 same scale.  The points in the image plane show the observed image
 positions, and the point in the source plane shows the inferred
 source position.  The value of $\mu$ gives the total magnification
 in each model.
\label{fig:clus-crit}}
\end{figure*}

We first consider models where the scale radius of the cluster is
fixed as $r_{\rm s} = 40''$ (we shall justify this choice below).
Figure~\ref{fig:clus-params} shows the allowed parameter ranges for
acceptable models.  First, panel (a) shows that the cluster
component is restricted to a fairly small (but not excessively
narrow) range of positions near the center of the image
configuration.  This is mainly a result of our upper limit on
the mass of the galaxy component; there is a certain enclosed
mass implied by the image separation, and if the galaxy cannot
contain all of that mass then the cluster component must lie
within the image configuration to make up the difference.  It is
interesting to note that even in these more complicated models
there still seems to be a small offset between the center of the
cluster component and the brightest cluster galaxy G1, although the
lower limit implied by our models is just $0\farcs71 = 3.6h^{-1}$kpc.

Figure~\ref{fig:clus-params}b shows that the allowed values for the
ellipticity and position angle of the galaxy G1 basically fill
the parameter space, so these parameters are not constrained by
the lens data.  We might want to impose an external constraint,
however.  Analyses of other lens systems show that the lensing mass is
typically aligned with the projected light distribution
\citep*{keeton98b,kochanek02}.  We may therefore prefer lens models
where the model galaxy is at least roughly aligned with the observed
galaxy, which has a position angle of $-19\fdg9$. To illustrate this
possible selection, we show all models but highlight those where the
position angle of the model galaxy is in the range $\theta_e =
-19\fdg9\pm20\fdg0$.  The broad $20\arcdeg$ uncertainties prevent this
constraint from being too strong.

Figure~\ref{fig:clus-params}c shows that there are some acceptable
models where the cluster potential is round, but most models have some
ellipticity that is aligned roughly North--South. This is
in good agreement with the distribution of member galaxies which is also
aligned roughly North--South (see Figure~\ref{fig:colorcut}). The
ellipticity $e \sim 0.2$--0.4 is actually quite large, considering that
this parameter describes the ellipticity of the potential, not that of
the density.  Figure~\ref{fig:clus-params}d shows that all of the
acceptable models require a fairly large tidal shear $\gamma \gtrsim
0.10$, and models where the galaxy is aligned with the observed galaxy
have a strong shear $\gamma \gtrsim 0.23$.  The shear tends to be
aligned East--West.  The fact that the models want both a large
cluster ellipticity and a large tidal shear strongly suggest that
there is complex structure in the cluster potential outside of
the image configuration.  It would be interesting to see whether
there is any evidence for such structure in, for example, X-rays
from the cluster.

Figure~\ref{fig:clus-crit} shows critical curves and caustics for
sample lens models.  The critical curves are not well determined.  The
outer, tangential critical curve can point either northeast (panel e) or
northwest (panel d), or it can have a 
complex shape (panel a).  Sometimes there is just one inner, radial
critical curve (panel e), but often there are two (panel c).  The
distance of the source from the caustic (and of the images from the
critical curve) varies from model to model, so the total magnification
can range from $\sim$50 to several hundred or even more.  Finally,
perhaps the most interesting qualitative result is that even the
image parities are not uniquely determined.  In most models (e.g.,
panels a--f) images A and D lie inside the critical curve and have
negative parity while B and C lie outside the critical curve and
have positive parity.  However, in some models (e.g., panels g--h)
the situation is reversed.  Having ambiguous image parities is very
rare in lens modeling.

So far we have only discussed models where the cluster has a
scale radius $r_{\rm s}=40''$.  We have also computed models with
$r_{\rm s} = (10, 20, 30, 50, 60)$ arcsec and we find that all of the
results are quite similar.  To understand what value of the scale
radius is reasonable, we must consider which (if any) of the models
have physically plausible cluster parameters.  Even though NFW models
are formally specified by two parameters $r_{\rm s}$ and $\kappa_{\rm
s}$, $N$-body simulations reveal that the two parameters are actually
correlated. NFW models therefore appear to form a one-parameter family
of models, although with some scatter which reflects the scatter of the
concentration parameter $c_{\rm vir}=r_{\rm vir}/r_{\rm s}$ ($r_{\rm
vir}$ is a virial radius of the cluster).
Figure~\ref{fig:NFWnorm} shows the predicted relation between
$r_{\rm s}$ and $\kappa_{\rm s}$, including the scatter.  For
comparison, it also shows the fitted values of $\kappa_{\rm s}$ in lens
models with different scale radii.  Models with $r_{\rm s}=10''$ or
$20''$ require $\kappa_{\rm s}$ much larger than expected,
corresponding to a halo that is too concentrated. Models with $r_{\rm s}
\ge 30''$, by contrast, overlap with the predictions and thus are
physically plausible.  We can therefore conclude very roughly that the
cluster component must have $r_{\rm s} \gtrsim 30''$ and a total
virial mass $M \gtrsim 10^{14}\,h^{-1}\,M_\odot$.

\vspace{0.5cm}
\centerline{{\vbox{\epsfxsize=7.7cm\epsfbox{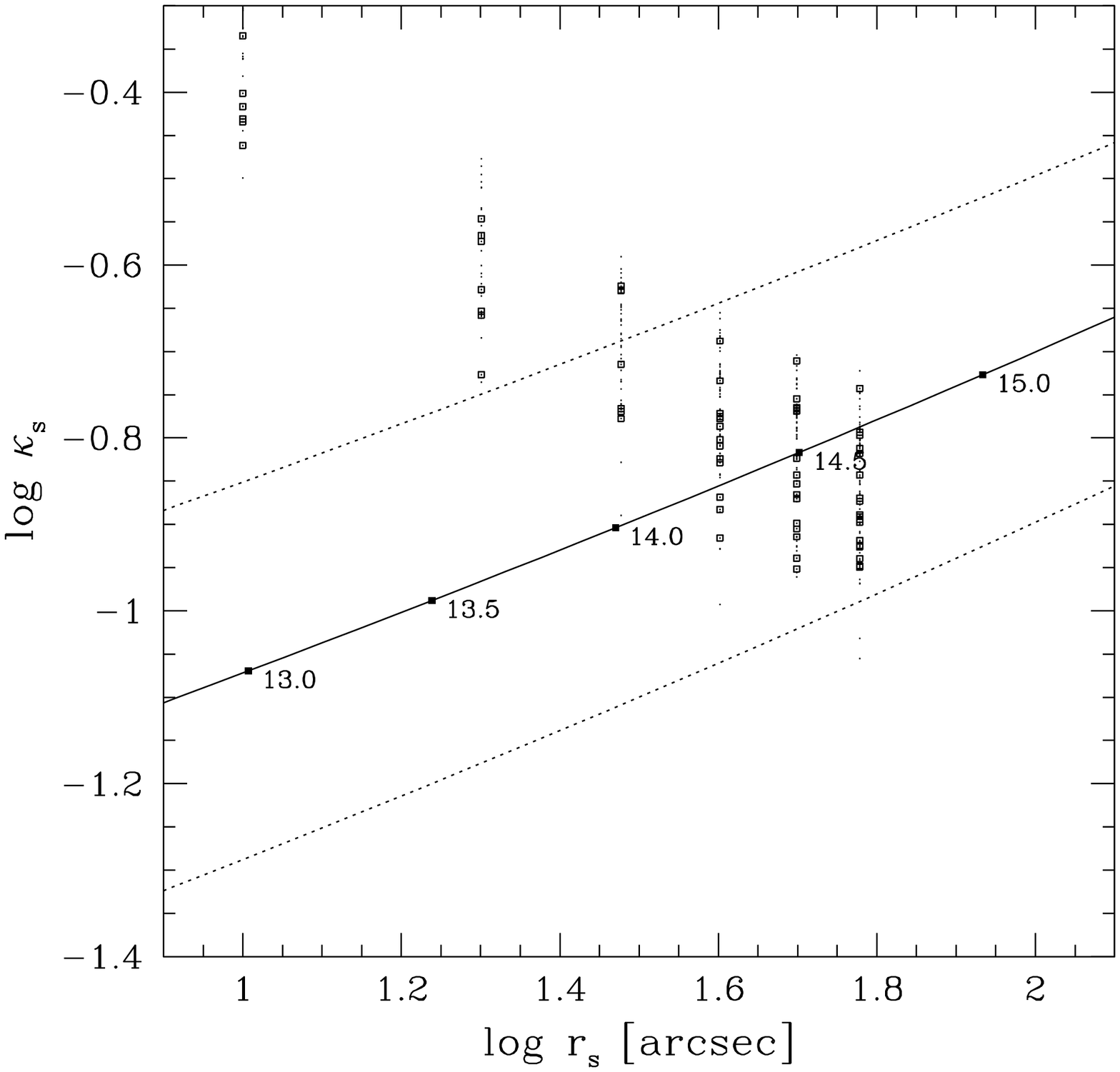}}}}
\figcaption{Relation between the cluster scale radius $r_{\rm s}$ and
 lensing strength $\kappa_{\rm s}$.  The solid line shows the predicted
 relation for clusters with the canonical median concentration, and the
 dotted lines show the 1$\sigma$ range due to the scatter in
 concentration (see \S\ref{sec:gnfw}). The labeled points show the
 value of $\log(M)$ (in units of $h^{-1}M_\odot$) at various points
 along the curve.  The points show fitted values of $\kappa_{\rm s}$ for
 lens models with $r_{\rm s}=(10,20,30,40,50,60)$ arcsec.  As in Figure
 \ref{fig:clus-params}, small points show all models, while boxes mark
 models where the model galaxy is roughly aligned with the observed
 galaxy.
\label{fig:NFWnorm}}
\vspace{0.5cm}

Finally, we can use the models to predict the time delays between
the images.  The models always predict that the time delay between
images C and D is the longest and the delay between A and B is the
shortest.  However, there is no robust prediction of the temporal
ordering: most models predict that the sequence should be C--B--A--D,
but a few models predict the reverse ordering D--A--B--C.  This is
a direct result of the ambiguity in the image parities, because the
leading image is always a positive-parity image
\citep*[e.g.,][]{schneider92}. We note, however, that all of the models
where the model galaxy is roughly aligned with the observed galaxy have
the C--B--A--D ordering.

Figure~\ref{fig:clus-tdel} shows the predictions for the long and short
time delays.  The long delay between C and D can be anything up to
$\sim\!3000\,h^{-1}$~days, while the short delay between A and B can
be up to $\sim\!37\,h^{-1}$~days.  For the models where the galaxy
is roughly aligned with the observed galaxy, the two delays are
approximately proportional to each other with
$\Delta t_{CD}/\Delta t_{BA} = 143\pm16$.  These results have
several important implications.  First, the A--B time delay should
be on the order of weeks or months, so it should be very feasible
to measure it, provided that the source has detectable variations.
Measuring the A--B delay would be very useful because it would
determine the temporal ordering, and thereby robustly determine the
image parities.  In addition, it would allow a good estimate of the
long C--D delay and indicate whether attempting to measure that delay
would be worthwhile.  Second, the enormous range of predicted time
delays means that constraining the Hubble constant with this system
\citep{refsdal64} will be difficult because of large systematic
uncertainties in the lens models.  Although \citet{koopmans03} recently
showed that it is possible to obtain useful constraints on the Hubble
constant even in a complex system with two mass components, the analysis
is very complex and requires extensive data including not just the image
positions and all of the time delays, but also an Einstein ring image
and the velocity dispersion of one of the mass components. Even if we
obtain such data for SDSS~J1004+4112 in the near future, it seems likely
that it will be difficult to obtain reliable constraints on the Hubble
constant given the complexity of the lens potential in SDSS~J1004+4112.
The time delays, however, would still be extremely useful, because they
would determine the temporal ordering and hence the image parities, and
they would provide constraints that can rule out many of the models
that are currently acceptable.

\vspace{0.5cm}
\centerline{{\vbox{\epsfxsize=7.7cm\epsfbox{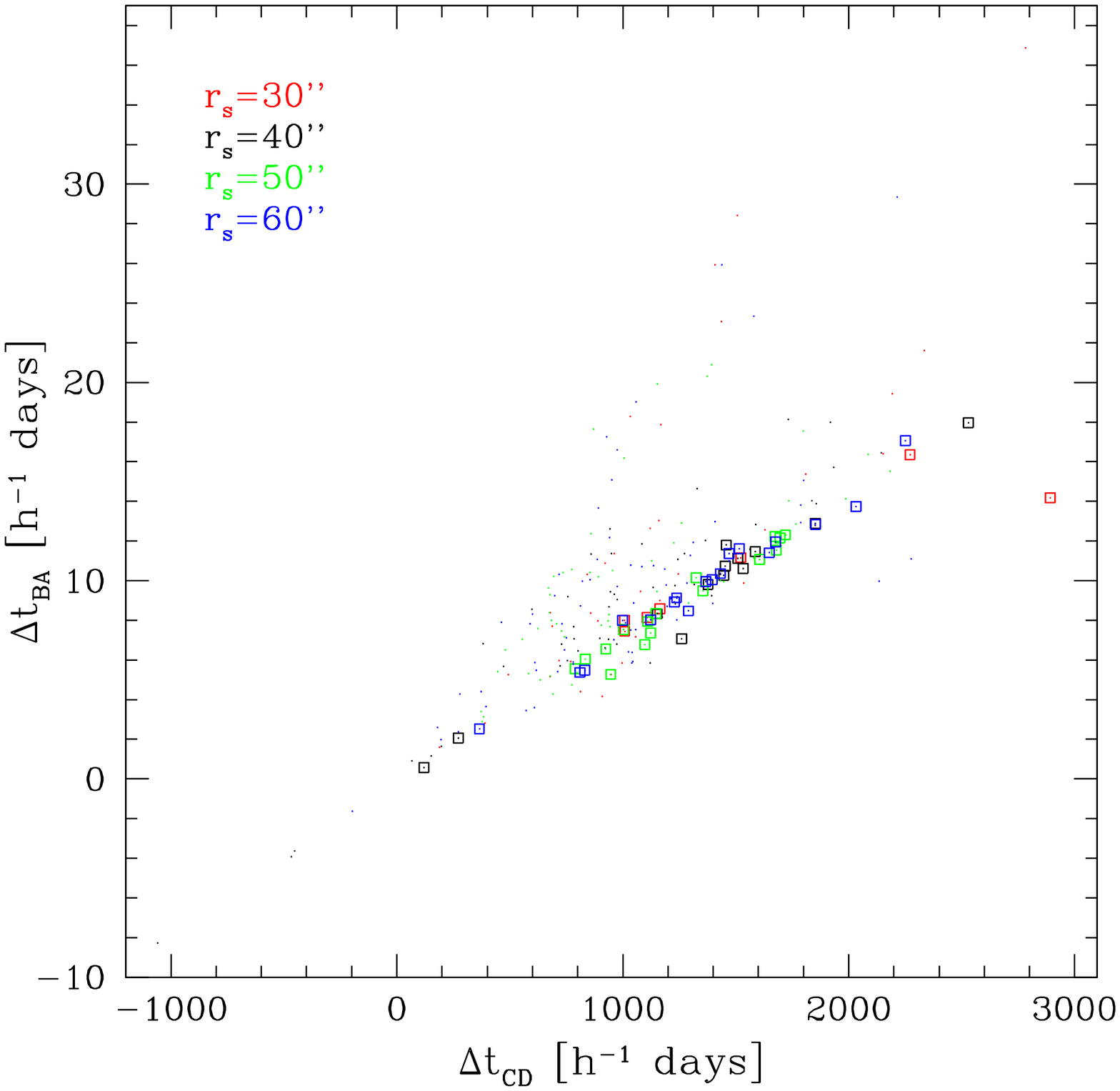}}}}
\figcaption{Predictions for the longest ($\Delta t_{CD}$) and shortest
($\Delta t_{BA}$) time delays, where $\Delta t_{ij}>0$ means image
$i$ leads image $j$, and vice versa.  Results are shown for models
where the cluster has scale length $r_{\rm s} = (30,40,50,60)$ arcsec.
As in Figure~\ref{fig:clus-params}, small points show all models,
while boxes mark models where the model galaxy is roughly aligned
with the observed galaxy.
\label{fig:clus-tdel}}
\vspace{0.5cm}

%%%%%%%%%%%%%%%%%%%%%%%%%%%%%%%%%%%%%%%%%%%%%%%%%%%%%%%%%%%
%%%%%%%%%%%%%%%%%%%%%%%%%%%%%%%%%%%%%%%%%%%%%%%%%%%%%%%%%%%
\section{Lens Statistics\label{sec:stat}}
%%%%%%%%%%%%%%%%%%%%%%%%%%%%%%%%%%%%%%%%%%%%%%%%%%%%%%%%%%%
%%%%%%%%%%%%%%%%%%%%%%%%%%%%%%%%%%%%%%%%%%%%%%%%%%%%%%%%%%%

In this section, we calculate the expected rate of large separation
lensing in the SDSS quasar sample.  The discovery of SDSS~J1004+4112
allows us to move past the upper limits obtained from previous
large separation lens searches, although at present the main thing
we can do is test whether the detection of one large separation lens
in the current sample is consistent with standard theoretical models
in the CDM scenario.

%%%%%%%%%%%%%%%%%%%%%%%%%%%%%%%%%%%%%%%%%%%%%%%%%%%%%%%%%%%
\subsection{Modeling Lens Probabilities}
%%%%%%%%%%%%%%%%%%%%%%%%%%%%%%%%%%%%%%%%%%%%%%%%%%%%%%%%%%%

We calculate lensing probabilities using spherical models for
simplicity.  Although halos in CDM simulations are in fact triaxial
\citep[e.g.,][]{jing02}, the spherical assumption is often adopted
in lens statistics calculations because deviations from spherical
symmetry mainly affect image multiplicities, not image separations
or the total optical depth for lensing
\citep[e.g.,][]{kochanek96,keeton97,chae03}.  While this result has
been obtained for isothermal lens potentials, checking it for more
general halos is beyond the scope of this paper and is the subject
of a follow-up analysis (M.~Oguri et al., in prep.).

%%%%%%%%%%%%%%%%%%%%%%%%%%%%%%%%%%%%%%%%%%%%%%%%%%%%%%%%%%%
\subsubsection{Lens Probabilities}
%%%%%%%%%%%%%%%%%%%%%%%%%%%%%%%%%%%%%%%%%%%%%%%%%%%%%%%%%%%

Let the physical image position in the lens plane and physical
source position in the source plane as $\xi$ and $\eta$,
respectively.  Consider the probability that a quasar at $z_{\rm S}$
with luminosity $L$ is strongly lensed. The probability of lensing
with image separation larger than $\theta$ is given by
%%%%%%%%%%%%%%%%%%%%%%%%%%%%%%%%%%%%%%%%%%%%%%%%%%%%%%%%%%%%%%%%%%%%%%%%
\begin{eqnarray}
P^{\rm B}(>\!\theta; z_{\rm S}, L) &=&
\int_{0}^{z_{\rm S}} dz_{\rm L}\,(1+z_{\rm L})^3\,\frac{c\,dt}{dz_{\rm L}}\nonumber\\
&&\times
\int_{M_{\rm min}}^{\infty} dM\, \frac{dn}{dM}\
\sigma_{\rm lens}\ B(z_{\rm S}, L)
\label{cpd_bias}
\end{eqnarray}
%%%%%%%%%%%%%%%%%%%%%%%%%%%%%%%%%%%%%%%%%%%%%%%%%%%%%%%%%%%%%%%%%%%%%%%%
where $\sigma_{\rm lens}=\pi\,\eta_{\rm r}^2\,D_{\rm OL}^2/D_{\rm OS}^2$
is the cross section for lensing, with $\eta_{\rm r}$ being the physical
radius of the radial caustic in the source plane.  The lower limit of
the mass integral is the mass $M_{\rm min}$ that corresponds to the
image separation $\theta$; this can be computed once the density
profile of the lens object is specified. The magnification bias
$B(z_{\rm S}, L)$ is \citep{turner80,turner84} 
%%%%%%%%%%%%%%%%%%%%%%%%%%%%%%%%%%%%%%%%%%%%%%%%%%%%%%%%%%%%%%%%%%%%%%%%
\begin{equation}
B(z_{\rm S}, L) = \frac{2}{\eta_{\rm r}^2\,\Phi(z_{\rm S}, L)}
\int_{0}^{\eta_{\rm r}} d\eta\ \eta\ \Phi(z_{\rm S}, L/\mu(\eta))\,
\frac{1}{\mu(\eta)},
\label{biasfactor}
\end{equation}
%%%%%%%%%%%%%%%%%%%%%%%%%%%%%%%%%%%%%%%%%%%%%%%%%%%%%%%%%%%%%%%%%%%%%%%%
where $\Phi(z_{\rm S}, L)$ is the luminosity function of source
quasars. Note that the magnification factor $\mu(\eta)$ may be
interpreted as the total magnification or the magnification of the
brighter or fainter image, depending on the observational selection
criteria \citep{sasaki93,cen94}. In this paper, we adopt the
magnification of the fainter image, because we concentrate on the
large separation lenses for which the images are completely deblended.

%%%%%%%%%%%%%%%%%%%%%%%%%%%%%%%%%%%%%%%%%%%%%%%%%%%%%%%%%%%
\subsubsection{Generalized NFW Profile\label{sec:gnfw}}
%%%%%%%%%%%%%%%%%%%%%%%%%%%%%%%%%%%%%%%%%%%%%%%%%%%%%%%%%%%

The lensing probability distribution at large separation reflects the
properties of dark halos, rather than galaxies
\citep{keeton01c,takahashi01,li02,oguri02b}.
For the statistics calculation, the debate over the inner slope of
the density profile seen in $N$-body simulations leads us to consider
the generalized version \citep{zhao96,jing00a} of the NFW density
profile (eq.~[\ref{nfw}]):
%%%%%%%%%%%%%%%%%%%%%%%%%%%%%%%%%%%%%%%%%%%%%%%%%%%%%%%%%%%%%%%%%%%%%%%%
\begin{equation}
 \rho(r)=\frac{\rho_{\rm crit}(z)\delta_{\rm c}(z)}
{\left(r/r_{\rm s}\right)^\alpha\left(1+r/r_{\rm s}\right)^{3-\alpha}}.
\label{gnfw}
\end{equation}
%%%%%%%%%%%%%%%%%%%%%%%%%%%%%%%%%%%%%%%%%%%%%%%%%%%%%%%%%%%%%%%%%%%%%%%%
While the correct value of $\alpha$ is still unclear, the existence of
cusps with $1\lesssim\alpha\lesssim1.5$ has been established in recent
$N$-body simulations \citep{navarro96,navarro97,moore99,ghigna00,jing00a,
klypin01,fukushige97,fukushige01,fukushige03a,power03,fukushige04,hayashi04}.
The case $\alpha=1$ corresponds to the original NFW profile, while the
case $\alpha=1.5$ resembles the profile proposed by \citet{moore99}.
The scale radius $r_{\rm s}$ is related to the concentration parameter
as $c_{\rm vir}=r_{\rm vir}/r_{\rm s}$.  Then the characteristic density
$\delta_{\rm c}(z)$ is given in terms of the concentration parameter:
%%%%%%%%%%%%%%%%%%%%%%%%%%%%%%%%%%%%%%%%%%%%%%%%%%%%%%%%%%%%%%%
\begin{equation}
\delta_{\rm c}(z)=\frac{\Delta_{\rm vir}(z)\Omega(z)}{3}
\frac{c_{\rm vir}^3}{m(c_{\rm vir})},
\end{equation}
%%%%%%%%%%%%%%%%%%%%%%%%%%%%%%%%%%%%%%%%%%%%%%%%%%%%%%%%%%%%%%%
where $m(c_{\rm vir})$ is
%%%%%%%%%%%%%%%%%%%%%%%%%%%%%%%%%%%%%%%%%%%%%%%%%%%%%%%%%%%%%%%
\begin{equation}
 m(c_{\rm vir})=\frac{c_{\rm vir}^{3-\alpha}}{3-\alpha}
\, {}_2F_1\left(3-\alpha, 3-\alpha; 4-\alpha; -c_{\rm vir}\right),
\end{equation}
%%%%%%%%%%%%%%%%%%%%%%%%%%%%%%%%%%%%%%%%%%%%%%%%%%%%%%%%%%%%%%%
with ${}_2F_1\left(a, b; c; x\right)$ being the hypergeometric function
\citep[e.g.,][]{keeton01c}. The {\it mean\/} overdensity
$\Delta_{\rm vir}(z)$ can be computed using the nonlinear spherical
collapse model \citep[e.g.,][]{nakamura97}.

We define $\tilde{\xi}\equiv\xi/r_{\rm s}$ and $\tilde{\eta}\equiv\eta
D_{\rm OL}/r_{\rm s}D_{\rm OS}$. Then the lensing deflection angle
$\beta(\tilde{\xi})$ is related to the dark halo profile as follows:
%%%%%%%%%%%%%%%%%%%%%%%%%%%%%%%%%%%%%%%%%%%%%%%%%%%%%%%%%%%%%%%%%%%%%%%%
\begin{equation}
\beta(\tilde{\xi})=\frac{4\kappa_{\rm s}}{\tilde{\xi}}\int_0^\infty dz
\int_0^{\tilde{\xi}} dx \frac{x}{\left(\sqrt{x^2+z^2}\right)^\alpha
\left(1+\sqrt{x^2+z^2}\right)^{3-\alpha}}.
\end{equation}
%%%%%%%%%%%%%%%%%%%%%%%%%%%%%%%%%%%%%%%%%%%%%%%%%%%%%%%%%%%%%%%%%%%%%%%%
The lensing strength parameter $\kappa_{\rm s}$ was defined in
equation (\ref{kappas}). For sources inside the caustic
($\eta<\eta_{\rm r}$), the lens equation has three solutions
$\tilde{\xi}_1>\tilde{\xi}_2>\tilde{\xi}_3$, where image \#1 is on
the same side of the lens as the source and images \#2 and \#3 are
on the opposite side.\footnote{The third image is usually predicted
to be very faint, so in practice just two images are actually
observed.} The lens image separation is then
%%%%%%%%%%%%%%%%%%%%%%%%%%%%%%%%%%%%%%%%%%%%%%%%%%%%%%%%%%%%%%%%%%%%%%%%
\begin{equation}
 \theta=\frac{r_{\rm s}(\tilde{\xi}_1+\tilde{\xi}_2)}{D_{\rm OL}}\simeq\frac{2r_{\rm s}\tilde{\xi}_{\rm t}}{D_{\rm OL}},
\label{sep}
\end{equation}
%%%%%%%%%%%%%%%%%%%%%%%%%%%%%%%%%%%%%%%%%%%%%%%%%%%%%%%%%%%%%%%%%%%%%%%%
where $\tilde{\xi}_{\rm t}$ is a radius of the tangential critical
curve \citep{hinshaw87,oguri02a}. The magnification of the fainter
image may be approximated by \citep{oguri02a}
%%%%%%%%%%%%%%%%%%%%%%%%%%%%%%%%%%%%%%%%%%%%%%%%%%%%%%%%%%%%%%%%%%%%%%%%
\begin{equation}
 \mu_{\rm faint}(\eta)\simeq\frac{\tilde{\xi}_{\rm t}}{\tilde{\eta}(1-\beta'(\tilde{\xi}_{\rm t}))}.
\label{approxfaint}
\end{equation}
%%%%%%%%%%%%%%%%%%%%%%%%%%%%%%%%%%%%%%%%%%%%%%%%%%%%%%%%%%%%%%%%%%%%%%%%
These approximations are sufficiently accurate over the range of
interest here \citep[see][]{oguri02a}. Although we adopt a selection
criterion that the flux ratios should be smaller than $10:1$, this
condition does not affect our theoretical predictions because the
flux ratios of strong lensing by NFW halos are typically much smaller
than $10:1$ \citep{oguri02a,rusin02}.

The concentration parameter $c_{\rm vir}$ depends on a halo's mass
and redshift. Moreover, even halos with the same mass and redshift
show significant scatter in the concentration which reflects the
difference in formation epoch \citep{wechsler02}, and which is well
described by a log-normal distribution.  For the median of this
distribution, we adopt the mass and redshift dependence reported by
\citet{bullock01} as a canonical model:
%%%%%%%%%%%%%%%%%%%%%%%%%%%%%%%%%%%%%%%%%%%%%%%%%%%%%%%%%%%%%%%
\begin{equation}
 c_{\rm Bullock}(M, z)=\frac{10}{1+z}\left(\frac{M}{M_*(0)}\right)^{-0.13},
\label{conmed_bullock}
\end{equation}
%%%%%%%%%%%%%%%%%%%%%%%%%%%%%%%%%%%%%%%%%%%%%%%%%%%%%%%%%%%%%%%
where $M_*(z)$ is the mass collapsing at redshift $z$ (defined by
$\sigma_M(z) = \delta_{\rm c} \equiv 1.68$). To study uncertainties
related to the concentration distribution we also consider other mass
and redshift dependences, e.g.,
%%%%%%%%%%%%%%%%%%%%%%%%%%%%%%%%%%%%%%%%%%%%%%%%%%%%%%%%%%%%%%%
\begin{equation}
 c_{\rm CHM}(M, z)=10.3(1+z)^{-0.3}\left(\frac{M}{M_*(z)}\right)^{-0.24(1+z)^{-0.3}},
\label{conmed_chm}
\end{equation}
%%%%%%%%%%%%%%%%%%%%%%%%%%%%%%%%%%%%%%%%%%%%%%%%%%%%%%%%%%%%%%%
from \citet*{cooray00}, and
%%%%%%%%%%%%%%%%%%%%%%%%%%%
\begin{equation}
 c_{\rm JS}(M, z)=2.44\sqrt{\frac{\Delta_{\rm vir}(z_c)}{\Delta_{\rm vir}(z)}}\left(\frac{1+z_c}{1+z}\right)^{3/2},
\label{conmed_js}
\end{equation}
%%%%%%%%%%%%%%%%%%%%%%%%%%%
from \citet{jing02}, with $z_c$ being the collapse redshift of the
halo of mass $M_{\rm vir}$. Note that these relations were derived
under the assumption of $\alpha=1$. We can extend them to
$\alpha\neq 1$ by multiplying the concentration by a factor $2-\alpha$
\citep{keeton01c,jing02}.

The statistics of large separation lenses are highly sensitive to the
degree of scatter in the concentration \citep{keeton01c,wyithe01,kuhlen04}.
\citet[][see also \citealp{wechsler02}]{bullock01} found
$\sigma_c \sim 0.32$ in their simulations. \citet{jing00b} found a
smaller scatter $\sigma_c\sim 0.18$ among well relaxed halos, but
\citet{jing02} found $\sigma_c\sim 0.3$ if all halos are considered.
Therefore we use $\sigma_c=0.3$ throughout the paper.

%%%%%%%%%%%%%%%%%%%%%%%%%%%%%%%%%%%%%%%%%%%%%%%%%%%%%%%%%%%
\subsubsection{Mass Function}
%%%%%%%%%%%%%%%%%%%%%%%%%%%%%%%%%%%%%%%%%%%%%%%%%%%%%%%%%%%

For the comoving mass function of dark matter halos, unless otherwise
specified we adopt equation (B3) of \citet{jenkins01}:
%%%%%%%%%%%%%%%%%%%%%%%%%%%
\begin{equation}
 \frac{dn_{\rm Jenkins}}{dM}=A\frac{\Omega_M\rho_{\rm crit}(0)}{M}\frac{d\ln\sigma_M^{-1}}{dM}\exp\left(-|\ln\sigma_M^{-1}(z)+B|^\epsilon\right),
\label{mf_jenkins}
\end{equation}
%%%%%%%%%%%%%%%%%%%%%%%%%%%
where $A=0.301$, $B=0.64$, and $\epsilon=3.82$. We use the
approximation of $\sigma_M$ given by \citet{kitayama96}, and the
shape parameter presented by \citet{sugiyama95}. Note that this
mass function is given in terms of the mean overdensity
$\Delta_{\rm c}=180$ instead of $\Delta_{\rm vir}(z)$.
Therefore, the mass function should be converted correctly
\citep[e.g.,][]{komatsu02}.  To study uncertainties related to
the mass function we also consider two other possibilities:
the mass function derived in the Hubble volume simulations,
$dn_{\rm Evrard}/dM$, which is given by equation (\ref{mf_jenkins})
with $A=0.22$, $B=0.73$, $\epsilon=3.86$ in terms of the mean
overdensity $\Delta_{\rm c}=200/\Omega(z)$ \citep{evrard02}; and
the mass function given by \citet{sheth99}
%%%%%%%%%%%%%%%%%%%%%%%%%%%
\begin{eqnarray}
 \frac{dn_{\rm STW}}{dM}&=&A\frac{\Omega_M\rho_{\rm crit}(0)}{M}\left[1+\left(\frac{\sigma_M^2(z)}{a\delta_{\rm c}^2}\right)^p\right]\nonumber\\
&&\times\sqrt{\frac{2a}{\pi}}\frac{\delta_{\rm c}}{\sigma_M(z)}\frac{d\ln\sigma_M^{-1}}{dM}\exp\left(-\frac{a\delta_{\rm c}^2}{2\sigma_M^2(z)}\right),
\end{eqnarray}
%%%%%%%%%%%%%%%%%%%%%%%%%%%
with $A=0.29$, $a=0.66$, $p=0.33$ in terms of the mean overdensity
$\Delta_{\rm c}=180$ \citep{white02}.

%%%%%%%%%%%%%%%%%%%%%%%%%%%%%%%%%%%%%%%%%%%%%%%%%%%%%%%%%%%
\subsection{Number of Lensed Quasars in the SDSS\label{sec:sdsscalc}}
%%%%%%%%%%%%%%%%%%%%%%%%%%%%%%%%%%%%%%%%%%%%%%%%%%%%%%%%%%%

Because the lensing probability depends on the source redshift
and luminosity, we compute the predicted number of lenses in
redshift and luminosity bins and then sum the bins.  Specifically,
let $N(z_j,i^*_k)$ by the number of quasars in a redshift range
$z_j-\Delta z/2<z<z_j+\Delta z/2$ that have a magnitude in the
range $i^*_k-\Delta i^*/2<i^*<i^*_k+\Delta i^*/2$. Then the
predicted total number of lensed quasars is
%%%%%%%%%%%%%%%%%%%%%%%%%%%%%%%%%%%%%%%%%%%%%%%%%%%%%%%%%%%%%%%%%%%%%%%%
\begin{equation}
N_{\rm lens}(>\!\theta)=\sum_{z_j}\sum_{i^*_k} N(z_j,i^*_k)\,
P(>\!\theta; z_j, L(i^*_k)).
\end{equation}
%%%%%%%%%%%%%%%%%%%%%%%%%%%%%%%%%%%%%%%%%%%%%%%%%%%%%%%%%%%%%%%%%%%%%%%%
We adopt bins of width $\Delta z=0.1$ and $\Delta i^*=0.2$.
The quasar sample we used comprises 29811 quasars with mean redshift
$\langle z\rangle=1.45$ (see Figure~\ref{fig:z_qso}), and roughly
corresponds to a sample with magnitude limit $i^*=19.1$
\citep{richards02}. 

To calculate the $B$-band absolute luminosity $L(i^*)$ corresponding
to observed magnitude $i^*$, we must estimate the cross-filter
K-correction $K_{Bi}(z)$. The K-correction calculated from the
composite quasar spectrum created from the SDSS sample by
\citet{vandenberk01} is shown in Figure~\ref{fig:kcor}. As a
simplification, one might use the following approximation:
%%%%%%%%%%%%%%%%%%%%%%%%%%%%%%%%%%%%%%%%%%%%%%%%%%%%%%%%%%%%%%%%%%%%%%%%
\begin{equation}
 K_{Bi}(z)=-2.5(1-\alpha_{\rm s})\log(1+z)-2.5\alpha_{\rm s}\log\left(\frac{7500}{4400}\right)-0.12,
\label{kcor_approx}
\end{equation}
%%%%%%%%%%%%%%%%%%%%%%%%%%%%%%%%%%%%%%%%%%%%%%%%%%%%%%%%%%%%%%%%%%%%%%%%
where the offset 0.12 mainly arises from the difference between
$AB(4400)$ and $B$ magnitudes \citep[calculated assuming
$\alpha_{\rm s}=0.5$;][]{schmidt95}. Throughout the paper, however,
we use the K-correction directly calculated from composite quasar
spectrum.

\vspace{0.5cm}
\centerline{{\vbox{\epsfxsize=7.7cm\epsfbox{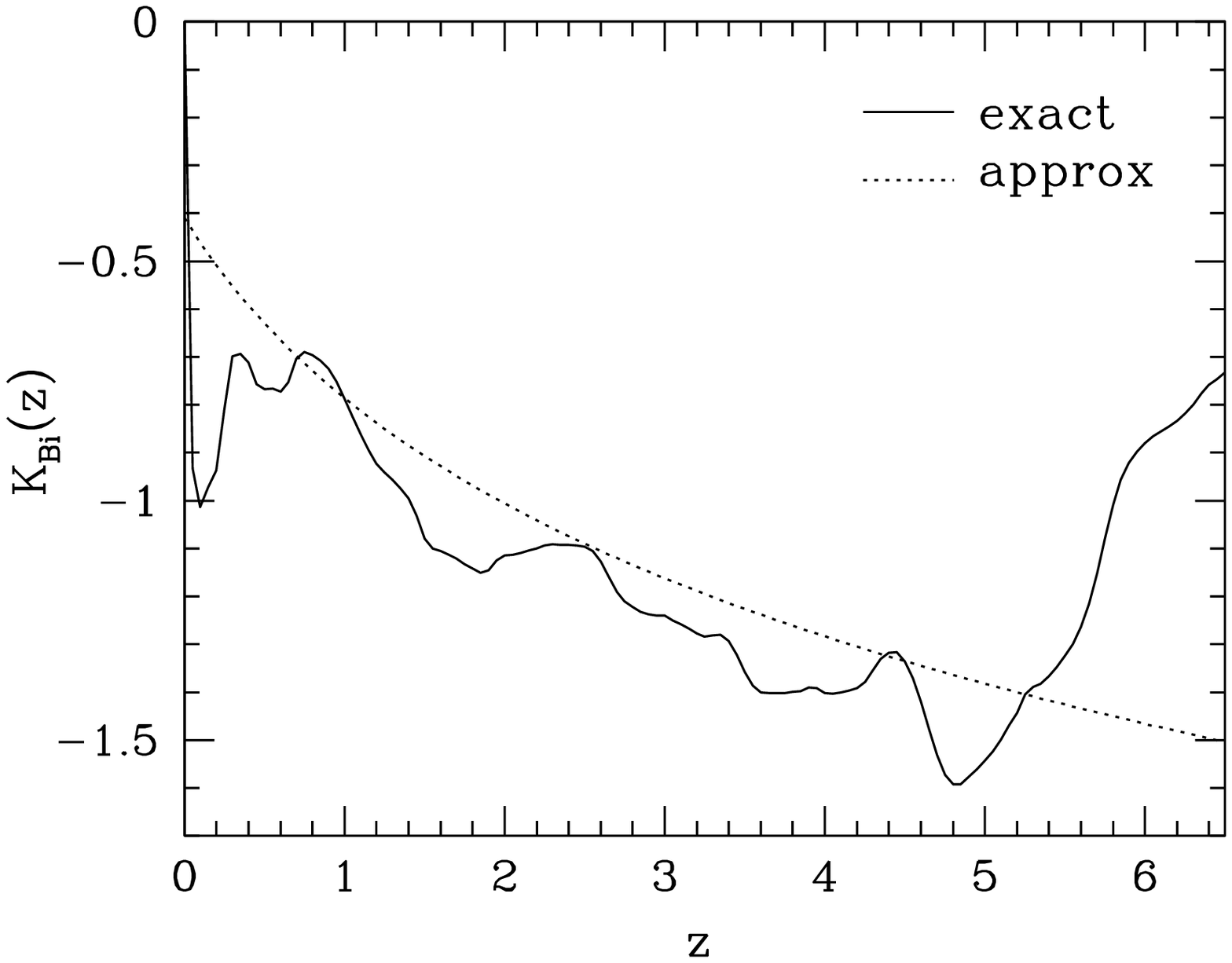}}}}
\figcaption{The cross-filter K-correction, computed from the SDSS composite
 quasar spectrum created by \citet{vandenberk01}. Dotted line indicate
 the approximation (eq.~[\ref{kcor_approx}]) with $\alpha_{\rm s}=0.5$.
\label{fig:kcor}}
\vspace{0.5cm}

The luminosity function of quasars is needed to compute magnification
bias. We adopt the standard double power law $B$-band luminosity
function \citep{boyle88}
%%%%%%%%%%%%%%%%%%%%%%%%%%%%%%%%%%%%%%%%%%%%%%%%%%%%%%%%%%%%%%%%%%%%%%%%
\begin{equation}
 \Phi(z_{\rm S},L)dL=\frac{\Phi_*}{[L/L_*(z_{\rm S})]^{\beta_l}+[L/L_*(z_{\rm S})]^{\beta_h}}\frac{dL}{L_*(z_{\rm S})}.
\end{equation}
%%%%%%%%%%%%%%%%%%%%%%%%%%%%%%%%%%%%%%%%%%%%%%%%%%%%%%%%%%%%%%%%%%%%%%%%
As a fiducial model of the evolution of the break luminosity, we assume
the form proposed by \citet{madau99},
%%%%%%%%%%%%%%%%%%%%%%%%%%%%%%%%%%%%%%%%%%%%%%%%%%%%%%%%%%%%%%%%%%%%%%%%
\begin{equation}
 L_*(z_{\rm S})=L_{*}(0)(1+z_{\rm S})^{\alpha_{\rm s}-1}\frac{e^{\zeta z_{\rm S}}(1+e^{\xi z_*})}{e^{\xi z_{\rm S}}+e^{\xi z_*}},
\end{equation}
%%%%%%%%%%%%%%%%%%%%%%%%%%%%%%%%%%%%%%%%%%%%%%%%%%%%%%%%%%%%%%%%%%%%%%%%
where a power-law spectral distribution for quasar spectrum has been
assumed, $f_\nu\propto \nu^{-\alpha_{\rm s}}$. \citet{wyithe02b}
determined the parameters so as to reproduce the low-redshift luminosity
function as well as the space density of high-redshift quasars for a
model with $\beta_h=3.43$ below $z_{\rm S}=3$, $\beta_h=2.58$ above
$z_{\rm S}=3$, and $\beta_l=1.64$. The resulting parameters are
$\Phi_*=624\,{\rm Gpc^{-3}}$, $L_{*}(0)=1.50\times 10^{11}~L_\odot$,
$z_*=1.60$, $\zeta=2.65$, and $\xi=3.30$. We call this model LF1.
To estimate the systematic effect, we also use another quasar
luminosity function (LF2) derived by \citet{boyle00}: $\beta_h=3.41$,
$\beta_l=1.58$ and an evolution of the break luminosity
$L_*(z_{\rm S})=L_*(0)10^{k_1z_{\rm S}+k_2z_{\rm S}^2}$ with
$k_1=1.36$, $k_2=-0.27$, and $M_*=-21.15+5\log h$.

%%%%%%%%%%%%%%%%%%%%%%%%%%%%%%%%%%%%%%%%%%%%%%%%%%%%%%%%%%%
\subsection{Results}
%%%%%%%%%%%%%%%%%%%%%%%%%%%%%%%%%%%%%%%%%%%%%%%%%%%%%%%%%%%

First we show the conditional probability distributions
%%%%%%%%%%%%%%%%%%%%%%%%%%%%%%%%%%%%%%%%%%%%%%%%%%%%%%%%%%%%%%%%%%%%%%%%
\begin{equation}
 \frac{dP}{dz_{\rm L}}(z_{\rm L}|\theta,z_{\rm S},L)\equiv\left|\frac{d^2P/dz_{\rm L}d\theta}{dP/d\theta}\right|,
\end{equation}
%%%%%%%%%%%%%%%%%%%%%%%%%%%%%%%%%%%%%%%%%%%%%%%%%%%%%%%%%%%%%%%%%%%%%%%%
%%%%%%%%%%%%%%%%%%%%%%%%%%%%%%%%%%%%%%%%%%%%%%%%%%%%%%%%%%%%%%%%%%%%%%%%
\begin{equation}
 \frac{dP}{d\ln M}(M|\theta,z_{\rm S},L)\equiv\left|\frac{d^2P/d\ln Md\theta}{dP/d\theta}\right|,
\end{equation}
%%%%%%%%%%%%%%%%%%%%%%%%%%%%%%%%%%%%%%%%%%%%%%%%%%%%%%%%%%%%%%%%%%%%%%%%
in order to identify the statistically reasonable ranges of redshift
and mass for the lensing cluster. Figure~\ref{fig:dist_z} shows the
conditional probability distribution of the lens redshift, and Figure
\ref{fig:dist_m} shows the conditional probability distribution of the
lens mass, given that the gravitational lens system SDSS~J1004+4112 has
image separation $\sim\!14''$, source redshift $z_{\rm S}=1.734$, and
apparent magnitude $i^*=18.86$. We find the most probable lens redshift
to be $z_{\rm L} \sim 0.5$, but the distribution is broad and the
measured redshift $z_{\rm L}=0.68$ is fully consistent with the
distribution. We also find a cluster mass
$M\sim 2$--$3\times 10^{14}\,h^{-1}\,M_\odot$ to be most probable for
this system. This result is in good agreement with the mass estimated
from the lens models (see Figure~\ref{fig:NFWnorm}).  Note that we do
not include information on the measured redshift $z_{\rm L}=0.68$ in
Figure~\ref{fig:dist_m}, which might cause a slight underestimate of
the lens mass.

\vspace{0.5cm}
\centerline{{\vbox{\epsfxsize=7.7cm\epsfbox{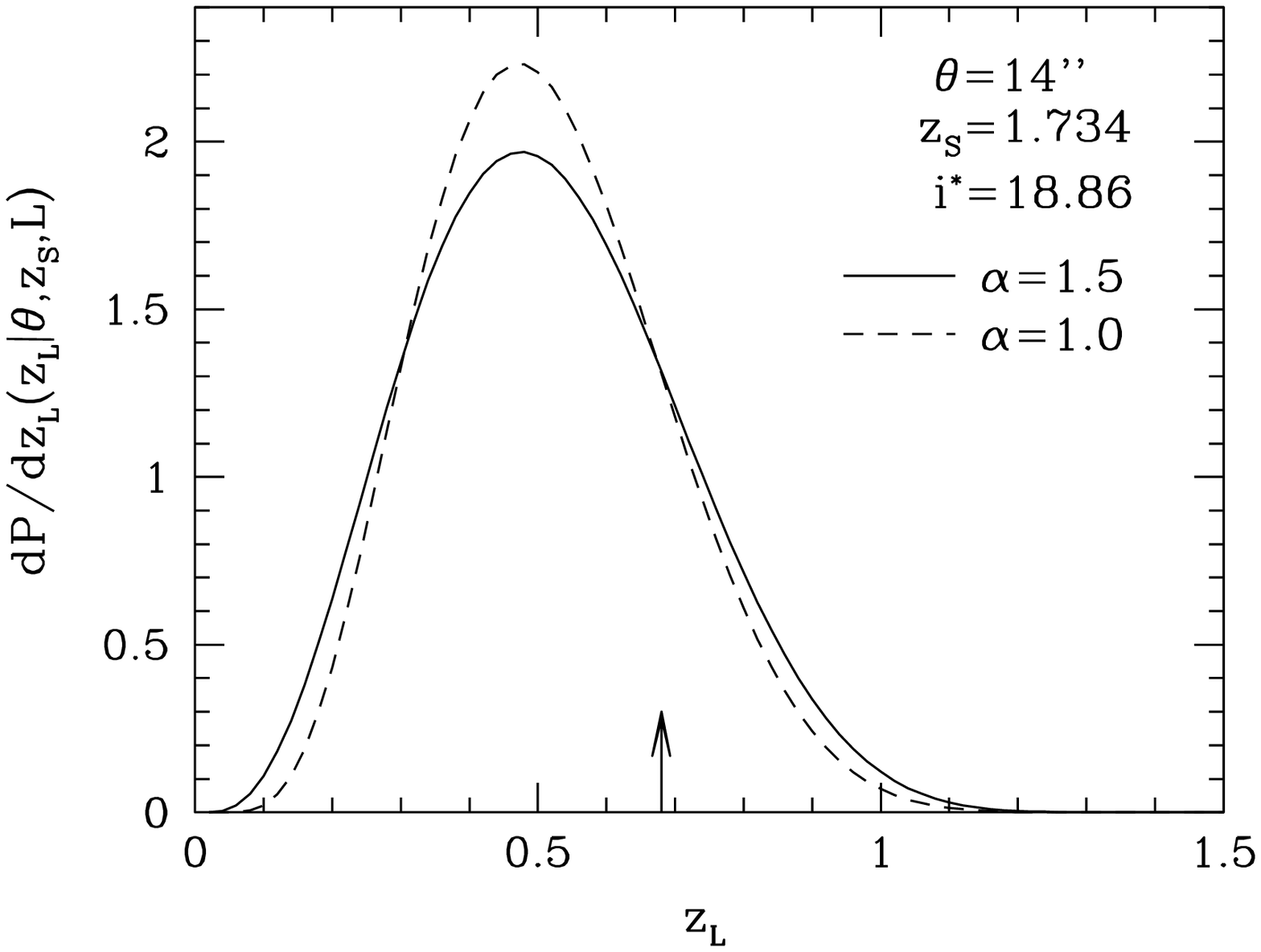}}}}
\figcaption{Conditional probability distributions for the lens redshift
 in SDSS~J1004+4112, given the image separation $\sim14''$, source
 redshift $z_{\rm S}=1.739$, and apparent magnitude $i^*=18.86$. Solid
 and dashed lines show the probability distributions with $\alpha=1.5$
 and $1.0$, respectively. The arrow shows the measured redshift of the
 lensing cluster. We assume $\sigma_8=1.0$.
 \label{fig:dist_z}}
\vspace{0.5cm}

\vspace{0.5cm}
\centerline{{\vbox{\epsfxsize=7.7cm\epsfbox{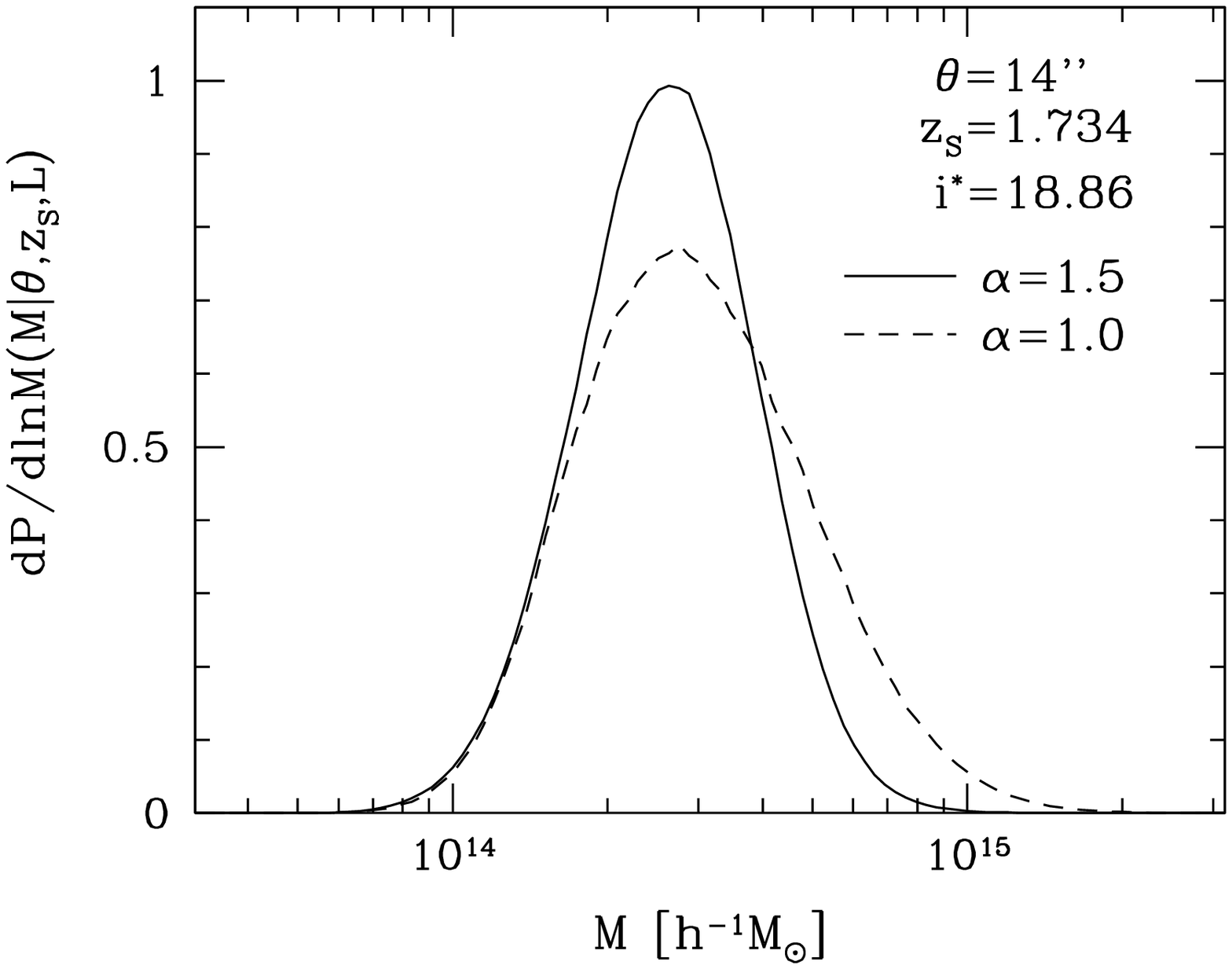}}}}
\figcaption{Same as Figure~\ref{fig:dist_z}, but the conditional
 probability distribution of the mass of the lens is plotted.
 \label{fig:dist_m}}
\vspace{0.5cm}

Next we consider the statistical implications of SDSS~J1004+4112.
Although our large separation lens search is still preliminary, and
we have several other candidates from the current SDSS sample that
still need follow-up observations, we can say that the current sample
contains {\it at least\/} one large separation lens system. This is
enough for useful constraints because of the complementary constraints
available from the lack of large separation lenses in previous lens
surveys.  Among the previous large separation lens surveys, we
adopt the CLASS $6''<\theta<15''$ survey comprising a statistically
complete sample of 9284 flat-spectrum radio sources \citep{phillips01b}.
For the CLASS sample, we use a source redshift $z_{\rm S}=1.3$
\citep{marlow00} and a flux distribution $N(S) dS \propto S^{-2.1} dS$
\citep{phillips01b} to compute the expected number of large separation
lenses.

\vspace{0.5cm}
\centerline{{\vbox{\epsfxsize=7.7cm\epsfbox{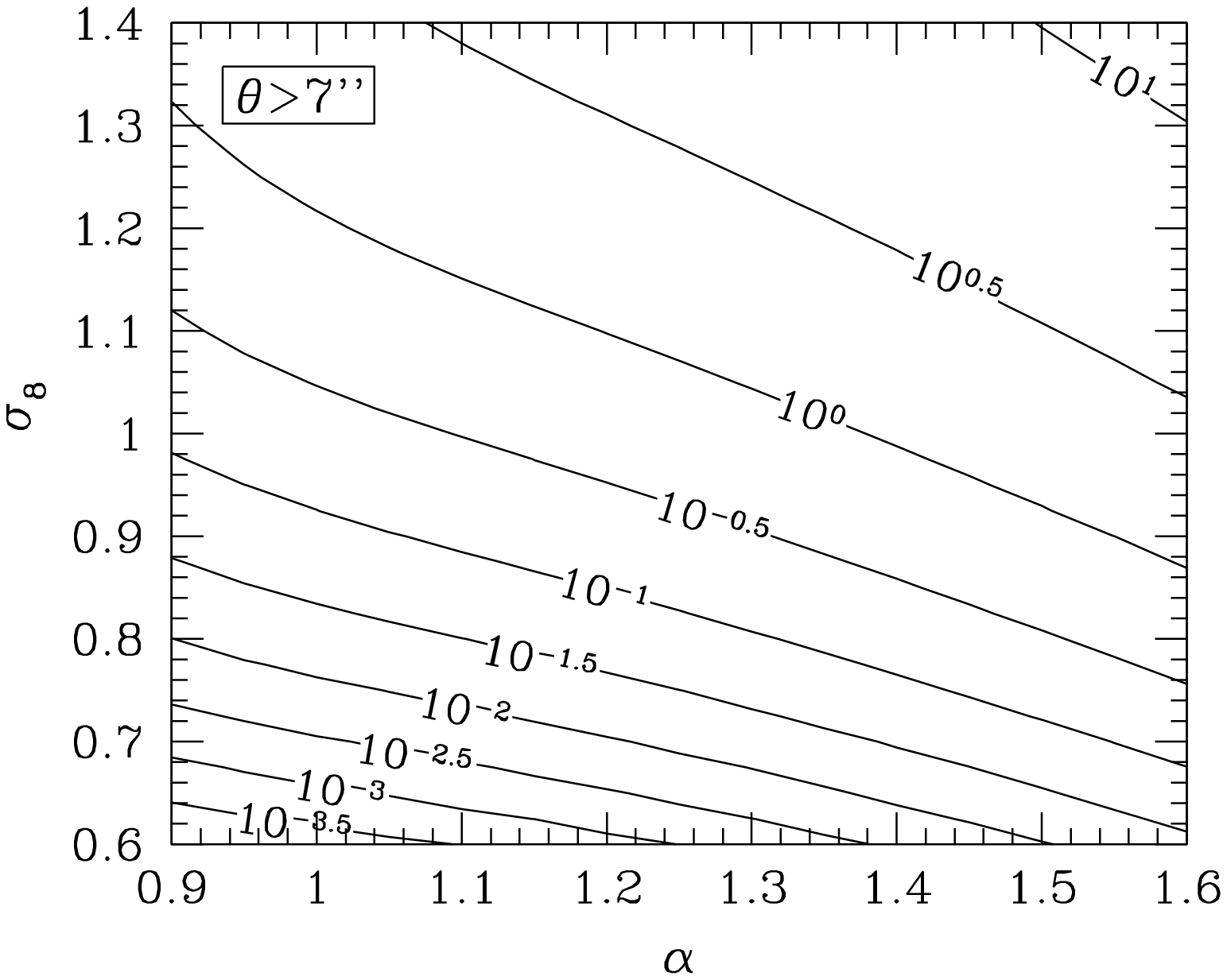}}}}
\figcaption{Contours of the predicted number of large separation
 ($\theta>7''$) lenses in the current SDSS sample in the
 $(\alpha,\sigma_8)$ plane.
\label{fig:cont}}
\vspace{0.5cm}

\vspace{0.5cm}
\centerline{{\vbox{\epsfxsize=7.7cm\epsfbox{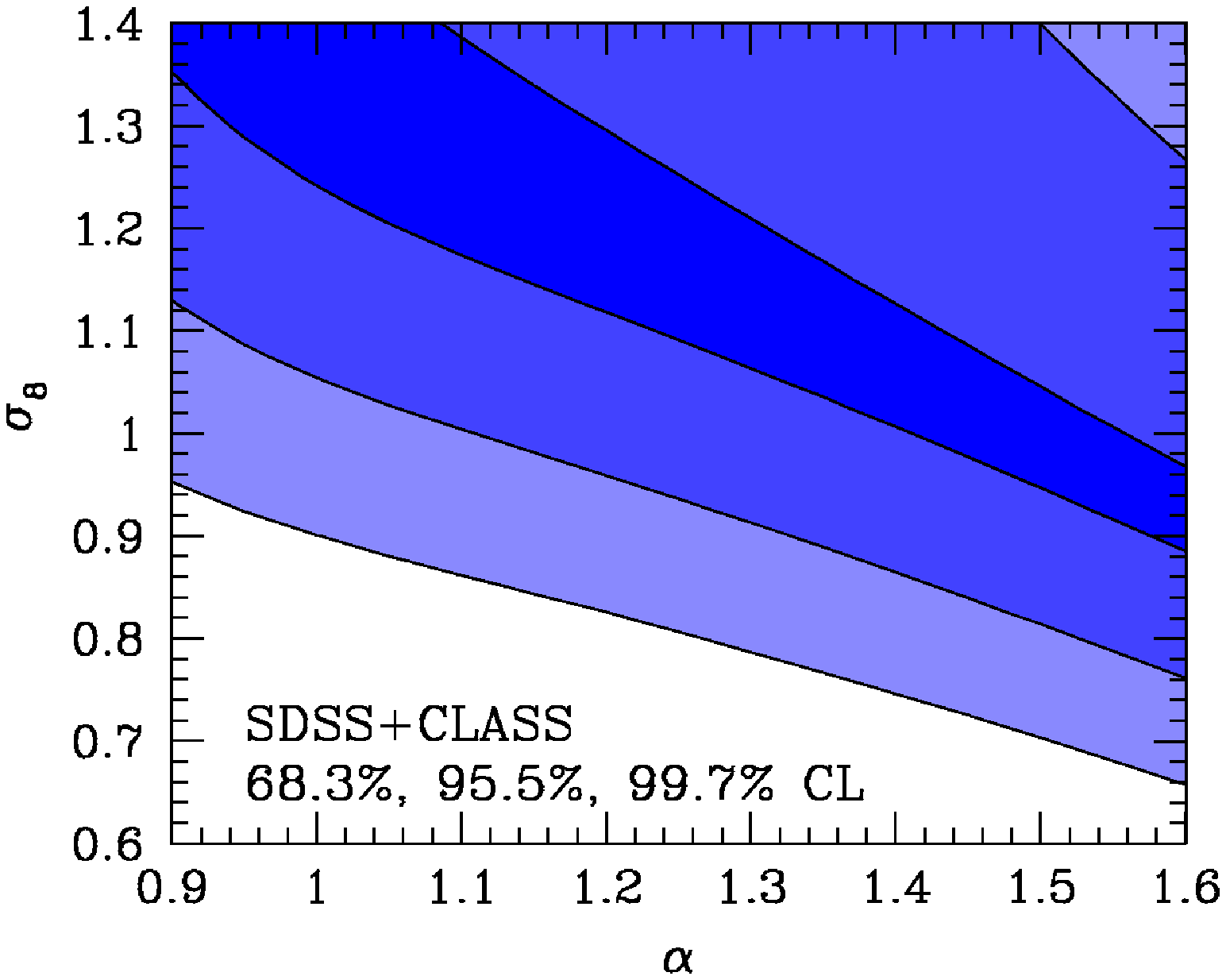}}}}
\figcaption{Constraints from both SDSS and CLASS in the $(\alpha,\sigma_8)$
 plane. The discovery of one large separation ($\theta>7''$) lens in SDSS
 provides lower limits on $\alpha$ and $\sigma_8$, while the lack of large
 separation lenses ($6''<\theta<15''$) in CLASS yields the upper limit.
 The regions in which both SDSS and CLASS limits are satisfied are shown
 by the shadings. The confidence levels are $68.3\%$, $95.5\%$, and
 $99.7\%$ in the dark, medium, and light shaded regions, respectively.
 \label{fig:limit}}
\vspace{0.5cm}

Figure~\ref{fig:cont} shows contours of the predicted number of large
separation lenses with $\theta>7''$ in the SDSS quasar sample.  Since
the number of lenses is very sensitive to both the inner slope of the
density profile $\alpha$ and the mass fluctuation normalization
$\sigma_8$ \citep[e.g.,][]{oguri02b}, we draw contours in the
$(\alpha,\sigma_8)$ plane. Constraints from the existence of
SDSS~J1004+4112 together with the lack of large separation lenses in
the CLASS sample are shown in Figure~\ref{fig:limit}. To explain both
observations, we need a relatively large $\alpha$ or $\sigma_8$, such
as $\sigma_8=1.0^{+0.4}_{-0.2}$ (95\% confidence) for $\alpha=1.5$.
This value is consistent with other observations such as cosmic
microwave background anisotropies \citep{spergel03}.  By contrast, if
we adopt $\alpha=1$ then the required value of $\sigma_8$ is quite
large, $\sigma_8 \gtrsim 1.1$.  Thus, our result might be interpreted
as implying that dark matter halos have cusps steeper than $\alpha=1$.
Alternatives to collisionless CDM, such as self-interacting dark matter
\citep{spergel00} or warm dark matter \citep*{bode01}, tend to produce
less concentrated mass distributions which are effectively expressed
by low $\alpha$; such models would fail to explain the discovery of
SDSS~J1004+4112 unless $\sigma_8$ is unexpectedly large. This result
is consistent with results from strong lensing of galaxies by clusters
(i.e., giant arcs), which also favors the collisionless CDM model
\citep*[][but see \citealt*{sand04} for different conclusion]{smith01,
meneghetti01,miralda02,gavazzi03,oguri03b,wambsganss04,dalal04}.
We note that the abundance of large separation lenses produces a
degeneracy between $\alpha$ and $\sigma_8$ seen in
Figure~\ref{fig:limit}, but additional statistics such as the
distribution of time delays can break the degeneracy \citep{oguri02a}.  

Table~\ref{table:lens} summarizes the sensitivity of our predictions
to various model parameters.  The uncertainties in our predictions
are no more than a factor of 2--3, dominated by uncertainties in the
concentration parameter and the matter density $\Omega_M$.  This error
roughly corresponds to $\Delta\sigma_8\sim 0.1$, and so does not
significantly change our main results. There may be larger systematic
errors associated with effects we have not considered in this paper.
For example, triaxiality in cluster halos is very important in arc
statistics because it can dramatically increase the length of the
tangential caustic that gives rise to giant arcs
\citep{oguri03b,meneghetti03a,dalal04}.  The effect would seem to be
less important in the statistics of lensed quasars, which depend
mainly on the area enclosed by the radial caustic, but it still needs
to be examined quantitatively.  The presence of a central galaxy is
thought to have a small impact on arc statistics \citep*{meneghetti03b},
but the complexity of the lens potentials we found in our mass modeling
suggests that the effect of the galaxy on the statistics of lensed
quasars also needs to be considered. 

%%%%%%%%%%%%%%%%%%%%%%%%%%%%%%%%%%%%%%%%%%%%%%%%%%%%%%%%%%%
%%%%%%%%%%%%%%%%%%%%%%%%%%%%%%%%%%%%%%%%%%%%%%%%%%%%%%%%%%%
\section{Summary and Conclusion\label{sec:discussion}}
%%%%%%%%%%%%%%%%%%%%%%%%%%%%%%%%%%%%%%%%%%%%%%%%%%%%%%%%%%%
%%%%%%%%%%%%%%%%%%%%%%%%%%%%%%%%%%%%%%%%%%%%%%%%%%%%%%%%%%%

We have presented the candidate selection and follow-up observations
of the first cluster-scale lensed quasar, SDSS~J1004+4112.  The system
consists of four components with image separation $\theta \sim 14''$,
and was selected from the large separation lens search in the SDSS.
The spectroscopic and photometric follow-up observations confirm
SDSS~J1004+4112 to be a lens system; spectroscopic observations of
four components showed that they have nearly identical spectra with
$z=1.734$. Deep images and spectroscopy of nearby galaxies indicate
that there is a cluster of galaxies with $z=0.68$, whose center is
likely to be among the four components.  We conclude that the cluster
is responsible for this large separation lens. Differences between
the \ion{C}{4} emission line profiles in the four images remain
puzzling, and it will be interesting to reobserve the profiles at
several epochs to search for variability that might explain the
differences.

We have shown that reasonable mass models can successfully reproduce
the observed properties of the lens.  When we consider models that
include both the cluster potential and the brightest cluster galaxy,
we find a broad range of acceptable models.  Despite the diversity
in the models, we find several general and interesting conclusions.
First, there appears to be a small ($\gtrsim\!4\,h^{-1}$~kpc) offset
between the brightest cluster galaxy and the center of the cluster
potential.  Such an offset is fairly common in clusters
\citep[e.g.,][]{postman95}.  Second, the cluster potential is
inferred to be elongated roughly North--South, which is consistent
with the observed distribution of apparent member galaxies.  Third,
we found that a significant external shear $\gamma\sim 0.2$ is needed
to fit the data, even when we allow the cluster potential to be
elliptical.  This may imply that the structure of the cluster potential
outside of the images is more complicated than simple elliptical
symmetry.  Fourth, given the broad range of acceptable models, we
cannot determine even the parities and temporal ordering of the images,
much less the amplitudes of the time delays between the images.
Measurements of any of the time delays would therefore provide powerful
new constraints on the models.  We note that the complexity of the
lens potential means that the time delays will be more useful for
constraining the mass model than for trying to measure the Hubble
constant.

Our modeling results suggest that further progress will require new
data (rather than refinements of current data).  The interesting
possibilities include catalogs of confirmed cluster members, X-ray
observations, and weak lensing maps, not to mention measurement of
time delays and confirmation of lensed arcs (either the possible
arclets we have identified, or others).  For instance, with an
estimated cluster mass of $M \sim 3\times 10^{14}\,h^{-1}\,M_\odot$,
the estimated X-ray bolometric flux is
$S_X \sim 10^{-13}\,{\rm erg\,s^{-1}cm^{-2}}$, which means that the
cluster should be accessible with the {\it Chandra} and
{\it XMM-Newton} X-ray observatories; the excellent spatial of
{\it Chandra} may be particularly useful for separating the diffuse
cluster component from the bright quasar images (which have a total
X-ray flux $S_X \sim 2\times 10^{-12}{\rm erg\,s^{-1}cm^{-2}}$ in
the {\it ROSAT} All Sky Survey).  The confirmation of lensed arclets
would be very valuable, as they would provide many more pixels'
worth of constraints on the complicated lens potential.  In
principle, mapping radio jets in the quasar images could unambiguously
reveal the image parities \citep[e.g.,][]{gorenstein88,garrett94},
but unfortunately the quasar appears to be radio quiet as it is not
detected in radio sky surveys such as the FIRST survey \citep*{becker95}. 

Although the large separation lens search in the SDSS is still
underway, we can already constrain model parameters from the
discovery of SDSS~J1004+4112.  The existence of at least one large
separation lens in SDSS places a lower limit on the lensing
probability that complements the upper limits from previous surveys.
Both results can be explained if clusters have the density profiles
predicted in the collisionless CDM scenario and moderate values of
the mass fluctuation parameter $\sigma_8$.  In particular we find
$\sigma_8=1.0^{+0.4}_{-0.2}$ (95\% confidence) assuming the inner
density profile of dark matter halos has the form
$\rho \propto r^{-\alpha}$ with $\alpha=1.5$.  The value of $\sigma_8$
is, however, degenerate with $\alpha$ such that smaller values of
$\alpha$ require larger values of $\sigma_8$.  Various systematic
errors are estimated to be $\Delta\sigma_8 \sim 0.1$, dominated by
uncertainties in the distribution of the concentration parameter
$c_{\rm vir}$ and in the matter density parameter $\Omega_M$.
Other systematic effects, such as triaxiality in the cluster potential
and the presence of a central galaxy, remain to be considered.  Still,
our overall conclusion is that the discovery of SDSS~J1004+4112 is
fully consistent with the standard model of structure formation (i.e.,
CDM with $\sigma_8 \sim 1$).

In summary, SDSS~J1004+4112 is a fascinating new lens system that
illustrates how large separation lenses can be used to probe the
properties of clusters and test models of structure formation.  The
full SDSS sample is expected to contain several more large separation
lenses.  The complete sample of lenses, and the distribution of their
image separations, will be extremely useful for understanding the
assembly of structures from galaxies to clusters.  More immediately,
the discovery of a quasar lensed by a cluster of galaxies fulfills
long-established theoretical predictions and resolves uncertainties
left by previously unsuccessful searches.

%%%%%%%%%%%%%%%%%%%%%%%%%%%%%%%%%%%%%%%%%%%%%%%%%%%%%%%%%%%%%%%
%%%%%%%%%%%%%%%%%%%%%%%%%%%%%%%%%%%%%%%%%%%%%%%%%%%%%%%%%%%%%%%
\acknowledgments
%%%%%%%%%%%%%%%%%%%%%%%%%%%%%%%%%%%%%%%%%%%%%%%%%%%%%%%%%%%%%%%
%%%%%%%%%%%%%%%%%%%%%%%%%%%%%%%%%%%%%%%%%%%%%%%%%%%%%%%%%%%%%%%

We thank Don York for valuable comments on the manuscript, and an
anonymous referee for many useful suggestions. Part of the work reported
here was done at the Institute of Geophysics and Planetary Physics,
under the auspices of the U.S. Department of Energy by Lawrence
Livermore National Laboratory under contract No.~W-7405-Eng-48.
Funding for the creation and distribution of the SDSS archive has
been provided by the Alfred P.\ Sloan Foundation, the Participating
Institutions, the National Aeronautics and Space Administration, the
National Science Foundation, the U.S. Department of Energy, the
Japanese Monbukagakusho, and the Max Planck Society. The SDSS
Web site is http://www.sdss.org/.

The SDSS is managed by the Astrophysical Research Consortium
(ARC) for the Participating Institutions. The Participating
Institutions are The University of Chicago, Fermilab, the Institute
for Advanced Study, the Japan Participation Group, The Johns Hopkins
University, Los Alamos National Laboratory, the Max-Planck-Institute
for Astronomy (MPIA), the Max-Planck-Institute for Astrophysics (MPA),
New Mexico State University, the University of Pittsburgh, Princeton
University, the United States Naval Observatory, and the University
of Washington.

This work is based in part on data collected at Subaru Telescope,
which is operated by the National Astronomical Observatory of Japan.
Some of the Data presented herein were obtained at the W.\ M.\ Keck
Observatory, which is operated as a scientific partnership between
the California Institute of Technology, the University of California,
and the National Aeronautics and Space Administration. The Observatory
was made possible by the generous financial support of the W.\ M.\
Keck Foundation. This work is also based in part on observations
obtained with the Apache Point Observatory 3.5-meter telescope, which
is owned and operated by the Astrophysical Research Consortium.  We
thank the staffs of Subaru, Keck, and APO 3.5-meter telescopes for
their excellent assistance.
The authors wish to recognize and acknowledge the very significant
cultural role and reverence that the summit of Mauna Kea has always
had within the indigenous Hawaiian community.  We are most fortunate
to have the opportunity to conduct observations from this mountain.

%%%%%%%%%%%%%%%%%%%%%%%%%%%%%%%%%%%%%%%%%%%%%%%%%%%%%%%%%%%%%%%%%%%%%%%
%\clearpage
%%%%%%%%%%%%%%%%%%%%%%%%%%%%%%%%%%%%%%%%%%%%%%%%%%%%%%%%%%%%%%%%%%%%%%%

%%%%%%%%%%%%%%%%%%%%%%%%%%%%%%%%%%%%%%%%%%%%%%%%%%%%%%%%%%%%%%%%%%%%%%%
\clearpage
%%%%%%%%%%%%%%%%%%%%%%%%%%%%%%%%%%%%%%%%%%%%%%%%%%%%%%%%%%%%%%%%%%%%%%%

\begin{deluxetable}{crrrrrr}
\tablewidth{0pt}
 \tablecaption{Photometry of SDSS~J1004+4112\label{table:sdssphoto}}
\tablehead{ \colhead{Object} & \colhead{$i^*$} & \colhead{$u^*-g^*$} &
 \colhead{$g^*-r^*$} & \colhead{$r^*-i^*$} & \colhead{$i^*-z^*$} &
 \colhead{Redshift}}
\startdata
A & $18.46\pm0.02$ & $0.15\pm0.05$ & $-0.03\pm0.04$ &
 $0.24\pm0.03$ & $0.02\pm0.05$ & $1.7339\pm0.0001$\\
B & $18.86\pm0.06$ & $0.18\pm0.08$ & $-0.05\pm0.08$ &
 $0.23\pm0.08$ & $-0.03\pm0.09$ & $1.7335\pm0.0001$\\
C & $19.36\pm0.03$ & $0.03\pm0.05$ & $-0.03\pm0.04$ &
 $0.38\pm0.04$ & $0.05\pm0.08$ & $1.7341\pm0.0002$\\
D & $20.05\pm0.04$ & $0.15\pm0.09$ & $0.15\pm0.05$&
 $0.46\pm0.05$ & $0.09\pm0.13$ & $1.7334\pm0.0003$\\
\enddata
\tablecomments{Magnitudes and colors for the four quasar images,
taken from the SDSS photometric data. Redshifts are derived from
Ly$\alpha$ lines in the Keck LRIS spectra (see Figure~\ref{fig:spec}). }
\end{deluxetable}

%%%%%%%%%%%%%%%%%%%%%%%%%%%%%%%%%%%%%%%%%%%%%%%%%%%%%%%%%%%%%%%%%%%%%%%

\begin{deluxetable}{cccrrc}
\tablewidth{0pt}
 \tablecaption{Astrometry of SDSS~J1004+4112\label{table:astrometry}}
\tablehead{ \colhead{Object} &
 \colhead{R.A.(J2000)} &
 \colhead{Dec.(J2000)} &
 \colhead{$\Delta$R.A.[arcsec]\tablenotemark{a}} &
 \colhead{$\Delta$Dec.[arcsec]\tablenotemark{a}}}
\startdata
A & 10 04 34.794 & $+$41 12 39.29 & $0.000\pm0.012$   & $0.000\pm0.012$ \\
B & 10 04 34.910 & $+$41 12 42.79 & $1.301\pm0.011$   & $3.500\pm0.011$\\
C & 10 04 33.823 & $+$41 12 34.82 & $-10.961\pm0.012$ & $-4.466\pm0.012$\\
D & 10 04 34.056 & $+$41 12 48.95 & $-8.329\pm0.007$  & $9.668\pm0.007$\\
G1   & 10 04 34.170 & $+$41 12 43.66 & $-7.047\pm0.053$  & $4.374\pm0.053$\\
\enddata
\tablecomments{Astrometry from the deep imaging data taken with
 Suprime-Cam (see \S\ref{sec:imaging}). The absolute coordinates are
 calibrated using the SDSS data.}
\tablenotetext{a}{Positions relative to component A.}
\end{deluxetable}

%%%%%%%%%%%%%%%%%%%%%%%%%%%%%%%%%%%%%%%%%%%%%%%%%%%%%%%%%%%%%%%%%%%%%%%

\begin{deluxetable}{ccc}
\tablewidth{0pt}
 \tablecaption{Subaru observations\label{table:subaruobs}}
\tablehead{ \colhead{Band} &
 \colhead{Exptime} &
 \colhead{$m_{\rm lim}$\tablenotemark{a}}}
\startdata
$g$ & 810  & 27.0 \\
$r$ & 1210 & 26.9 \\
$i$ & 1340 & 26.2 \\
$z$ & 180  & 24.0 \\
\enddata
\tablecomments{Total exposure time in seconds (exptime) and limiting
magnitude ($m_{\rm lim}$) for the Subaru deep imaging observations.}
\tablenotetext{a}{Defined by $S/N>5$ for point sources.}
\end{deluxetable}

%%%%%%%%%%%%%%%%%%%%%%%%%%%%%%%%%%%%%%%%%%%%%%%%%%%%%%%%%%%%%%%%%%%%%%%

\begin{deluxetable}{crrcc}
\tablewidth{0pt}
 \tablecaption{Constraints on mass models \label{table:posflux}}
\tablehead{ \colhead{Object} &
 \colhead{$x$[arcsec]\tablenotemark{a}} &
 \colhead{$y$[arcsec]\tablenotemark{a}} &
 \colhead{Flux[arbitrary]\tablenotemark{b}}&
 \colhead{PA[deg]\tablenotemark{c}}}
\startdata
A  & $ 0.000\pm0.012$ & $ 0.000\pm0.012$ & $1.0  \pm0.2  $ & \nodata \\
B  & $-1.301\pm0.011$ & $ 3.500\pm0.011$ & $0.682\pm0.136$ & \nodata \\
C  & $10.961\pm0.012$ & $-4.466\pm0.012$ & $0.416\pm0.083$ & \nodata \\
D  & $ 8.329\pm0.007$ & $ 9.668\pm0.007$ & $0.195\pm0.039$ & \nodata \\
G1 & $ 7.047\pm0.053$ & $ 4.374\pm0.053$ &  \nodata        & $-19.9$ \\
\enddata
\tablecomments{Summary of positions, flux ratios, and position angles
 (PA) of SDSS~J1004+4112 used in the mass modeling.}
\tablenotetext{a}{The positive directions of $x$ and $y$ are defined by
 West and North, respectively.}
\tablenotetext{b}{Errors are broadened to 20\% to account for possible
 systematic effects. }
\tablenotetext{c}{Degrees measured East of North.}
\end{deluxetable}

%%%%%%%%%%%%%%%%%%%%%%%%%%%%%%%%%%%%%%%%%%%%%%%%%%%%%%%%%%%%%%%%%%%%%%%

\begin{deluxetable}{ccccc}
\tablewidth{0pt}
 \tablecaption{Systematic effects\label{table:lens}}
\tablehead{ \colhead{} &
 \multicolumn{4}{c}{$N_{\rm lens}(>7'')$ for $(\alpha,\sigma_8)$} \\
 \cline{2-5}
 \colhead{Models} &
 \colhead{$(1.0,0.7)$} &
 \colhead{$(1.5,0.7)$} &
 \colhead{$(1.0,1.1)$} &
 \colhead{$(1.5,1.1)$}}
\startdata
fiducial model & $0.0027$ & $0.071$ & $0.47$ & $3.0$ \\
$c_{\rm Bullock}\rightarrow c_{\rm CHM}$ & $0.00002$ & $0.0097$ & $0.16$ & $1.8$ \\
$c_{\rm Bullock}\rightarrow c_{\rm JS}$ & $0.00049$ & $0.065$ & $0.17$ & $2.4$ \\
$dn_{\rm Jenkins}/dM\rightarrow dn_{\rm Evrard}/dM$ & $0.0044$ & $0.11$ & $0.53$ & $3.2$ \\
$dn_{\rm Jenkins}/dM\rightarrow dn_{\rm STW}/dM$ & $0.00075$ & $0.026$ & $0.22$ & $1.7$ \\
LF1$\rightarrow$LF2 & $0.0024$ & $0.066$ & $0.42$ & $2.8$ \\
$\Omega_M=0.27\rightarrow 0.22$ &  $0.00066$ & $0.027$ & $0.24$ & $1.8$ \\
$\Omega_M=0.27\rightarrow 0.32$ & $0.0083$ & $0.15$ & $0.81$ & $4.6$ \\
\enddata
\tablecomments{Sensitivity of the predicted number of large separation
 lensed quasars in the SDSS quasar sample to various changes in the
 statistics calculations.}
\end{deluxetable}

\end{document}